\documentstyle[preprint,subeqn,eqsecnum,aps,epsfig,floats]{revtex}

\begin{document}

\newcommand{\be}{\begin{equation}}
\newcommand{\ee}{\end{equation}}
\newcommand{\ba}{\begin{eqnarray}}
\newcommand{\ea}{\end{eqnarray}}
\newcommand{\bc}{\begin{center}}
\newcommand{\ec}{\end{center}}
\newcommand{\vs}{\vspace*{3mm}}
\newcommand{\dis}{\displaystyle}
\newcommand{\bay}{\begin{array}{rcl}}
\newcommand{\eay}{\end{array}}
\def\RN{Reis\-sner-Nord\-str\"{o}m }
\def\rc{\rho_{\rm crit}}
\def\rl{\rho_\Lambda}
\def\rt{\rho_{\rm tot}}
\def\ie{{\it i.e.}}
\def\lp{\ell_{\rm Pl}}
\def\mp{m_{\rm Pl}}
\def\tp{t_{\rm Pl}}
\def\om{\Omega_{\rm M}}
\def\oa{\Omega_{\Lambda}}
\def\ot{\Omega_{\rm tot}}
\def\tla{\widetilde{\lambda}_\ast}
\def\tom{\widetilde{\omega}_\ast}

\textwidth16cm

\setlength{\oddsidemargin}{0cm}
\setlength{\jot}{0.3cm}

\begin{titlepage}

\renewcommand{\thefootnote}{\fnsymbol{footnote}}
\renewcommand{\baselinestretch}{1}

\begin{flushright}
 {INFNCT/4/01,  MZ-TH/01-18}\\
\end{flushright}

\begin{center}

{\Large \sc Cosmology of the Planck Era from a\\
Renormalization Group for \\
Quantum Gravity}                   
\vspace{5mm}

{\large
A. Bonanno\footnote{ \tt abo@ct.astro.it}} \\
\vspace{0.5cm}
\noindent
{\it Osservatorio Astrofisico,
Via S.Sofia 78, I-95123 Catania, Italy\\
INFN Sezione di Catania, Corso Italia 57, I-95129 Catania, Italy}

\vspace{1cm}
{\large
M. Reuter \footnote{\tt reuter@thep.physik.uni-mainz.de}}\\
\vspace{0.5cm}
\noindent
{\it Institut f\"ur Physik, Universit\"at Mainz\\
Staudingerweg 7, D-55099 Mainz, Germany}
\end{center}                                                       

\begin{abstract}
Homogeneous and isotropic cosmologies of the Planck era before the classical 
Einstein equations become valid are studied taking quantum gravitational effects into
account. The cosmological evolution equations are renormalization group improved by including the scale
dependence of Newton's constant and of the cosmological constant as it is given by the flow
equation of the effective average action for gravity. It is argued that the Planck regime can be treated
reliably in this framework because gravity is found to become asymptotically free at short distances. The
epoch immediately after the initial singularity of the Universe is described 
by an attractor solution of the improved equations which is a direct manifestation of an 
ultraviolet attractive renormalization group fixed point. It is shown that quantum 
gravity effects in the very early Universe might provide a resolution to the horizon and 
flatness problem of standard cosmology, and could generate a scale-free spectrum of primordial
density fluctuations.
\end{abstract}




\end{titlepage}

\baselineskip 18pt

\section{Introduction} 
Two of the most frequently discussed limitations of the cosmological standard model
are the flatness and the horizon problem, respectively. These so-called 
``problems'' actually do not endanger the internal consistency of the standard 
model in the domain where it is  applicable but rather express the fact that in order to
describe the Universe as we observe it today the standard Friedmann-Robertson-Walker
evolution has to start from a set of highly non-generic initial conditions.
Typically these conditions are imposed at some time after the Planck era where the
classical Friedmann equations are supposed to become valid. The matter density $\rho$
of the present Universe is very close to the critical density $\rho_{\rm crit}$.
According to the evolution equations of the standard model this implies that the 
initial value for $\rho$ must have been finetuned to the critical density
with the enormous precision of about 60 decimal places if the initial conditions
are imposed at the Planck time. This phenomenon is referred to as the flatness
problem because a generic initial value for the density would never have led to the
large and almost flat Universe we observe today. More generally, if one allows for a
cosmological constant $\Lambda$, it is the total density $\rt = \rho +\rl$ with the 
vacuum energy density $\rl \equiv \Lambda / 8 \pi G$ which should be equal to $\rc$. 

A similar naturalness problem is posed by the high degree of isotropy of the cosmic 
microwave background radiation. From the observations we know that even those points
on the last scattering hypersurface which, according to the metric of the cosmological 
standard model, never have been in causal contact emit radiation at a temperature which 
is constant with a precision of about $10^{-4}$. Again, when equipped with 
sufficiently symmetric initial conditions the cosmological standard model
can describe the later evolution of such a highly isotropic universe, but clearly it
would be very desirable to identify some causal mechanism which explains why one
must start the classical evolution with these very special initial conditions. This is
usually called the horizon problem because those Robertson-Walker spacetimes which solve
the Friedmann equations have a particle horizon. Because of this horizon, there are
points on the last scattering surface whose backward light cones never intersect
and which are causally disconnected therefore. 
 
However, strictly speaking this is a ``problem'' only if one applies the standard model
in a domain where it is actually believed not to be valid any more. Whether or not a 
Robertson-Walker spacetime has a particle horizon depends only on the behavior
of its scale factor $a(t)$ in the limit $t\rightarrow 0$. In the ordinary radiation dominated
Universe we have $a\propto t^{1/2}$ which does lead to a horizon. However, we expect
that for the cosmological time $t$ very close to the big bang $(t=0)$ this behavior of
$a(t)$ will get modified by some sort of ``new physics''. If, say, $a\propto t^{\alpha}$
with $\alpha \geq 1$ during the very early evolution of the Universe then there is no
particle horizon. It might be that a causal mechanism which is operative during this 
early epoch, before the standard model becomes valid, can explain the observed
isotropy of the Universe. 

It is well known that the above naturalness problems can be addressed and, in a sense,
solved within the framework of inflationary cosmology \cite{brandinf}, for instance.
In the present paper we are going to propose a different physical mechanism which also
could lead to a solution of the horizon and the flatness problem.
Using renormalization group techniques we determine the leading quantum 
gravity corrections which modify the standard Friedmann-Robertson-Walker (FRW) cosmology
during the first few Planck times after the big bang. Within a certain approximation, 
which we shall describe in detail below, we find that immediately after the big bang
there is a period during which the scale factor increases linearly with time,
$a\sim t$. This means that the spacetime has no particle horizon. We shall set up 
a system of quantum corrected cosmological evolution equations for
$a(t)$, $\rho(t)$, $p(t)$ and for the now time dependent Newton constant and cosmological
constant. We shall argue that, because of a specific form of asymptotic freedom
enjoyed by quantum gravity, those equations are reliable even for times infinitesimally close
to the big bang where the gravitational coupling constant goes to zero. During the epoch 
directly after the big bang the quantum corrected equations are uniquely solved
by an essentially universal attractor-type solution. For a spatially flat geometry 
the attractor satisfies $\rho=\rl=\rt /2$ and $\rt=\rc$. For $t$ much larger than the 
Planck time, the quantum corrected solutions approach those of classical FRW 
cosmology. Since the quantum solutions are valid for all $t>0$, they automatically 
prepare the initial condition $\rt=\rc$ for the classical regime if one 
decides for a spatially flat Universe. Hence no finetuning is necessary.

In this paper we employ the Exact Renormalization Group approach to quantum gravity which 
has been developed in ref.\cite{mr}. Its basic ingredient is the effective average action
$\Gamma_k[g_{\mu\nu}]$, a Wilsonian coarse grained free energy which depends on a momentum
scale $k$. Loosely speaking, $\Gamma_k$ describes the dynamics of metrics which have been
averaged over spacetime volumes of linear dimension $k^{-1}$, {\it i.e.} $k$ is a measure for 
the resolution of the ``microscope'' with which a system is observed.  The functional 
$\Gamma_k[g_{\mu\nu}]$ defines an effective field theory appropriate for the scale $k$.
This means that, when evaluated at {\it tree} level, $\Gamma_k$ correctly describes all
gravitational phenomena, {\it including all loop effects}, if the typical momenta involved
are all of the order of $k$. The action $\Gamma_k$ is constructed in a similar way as the
ordinary effective action $\Gamma$ to which it reduces in the limit $k\rightarrow 0$.
It has the additional feature of a built-in infrared (IR) cutoff at the momentum $k$.
Quantum fluctuations with momenta $p^2>k^2$ are integrated out in the usual way, while the 
contributions coming from large-distance metric fluctuations with $p^2<k^2$ are not included in
$\Gamma_k$. When regarded as a function of $k$, $\Gamma_k$ describes a renormalization
group (RG) trajectory in the space of all action functionals. This trajectory can be 
determined by solving an exact functional renormalization group equation or 
``flow equation''. The trajectory interpolates between the classical action 
$S=\Gamma_{k\rightarrow \infty }$ and the ordinary effective action 
$\Gamma=\Gamma_{k\rightarrow 0}$. More precisely, in order to quantize a renormalizable 
fundamental theory with action $S$ one integrates the RG equation from an initial 
point $\Gamma_{\widehat{k}}=S$ down to $\Gamma_0 \equiv \Gamma$.  After appropriate renormalizations
one then lets ${\widehat k}\rightarrow \infty $. The RG equation can also be used in 
order to further evolve (coarse-grain) effective field theory actions from one scale to another.
In this case no UV limit $\widehat{k}\rightarrow\infty$ needs to be taken. 
The evolution of the effective average action from $k_1$ down to $k_2< k_1$ is always
well defined even if (as in the case at hand) the model defined by $\Gamma_1$ is not
perturbatively renormalizable.

Approximate yet nonperturbative solutions to the RG equation which do not require an expansion
in a small coupling constant can be obtained by the method of ``truncations''. The idea is to
project the RG flow from the infinite dimensional space of all action functionals onto
some finite dimensional subspace which is particularly relevant. In this manner the functional
RG equation becomes a system of ordinary differential equations for a finite set of generalized
coupling constants which serve as coordinates on this subspace. In ref. \cite{mr} the flow
was projected on the 2-dimensional space spanned by the operators $\int \sqrt{g}R$ and
$\int \sqrt{g}$ (``Einstein-Hilbert truncation''). The corresponding generalized couplings
are the scale dependent (``running'') Newton constant $G(k)$ and the cosmological
constant $\Lambda(k)$. In the original paper \cite{mr} the differential equations
governing the $k$-dependence of $G(k)$ and $\Lambda(k)$ were derived, and in
\cite{souma,bh2} their solutions were discussed further. In particular one finds that if
one increases $k$ from small values (large distances) to higher values (small
distances) the value of $G(k)$ decreases, {\it i.e.} gravity is asymptotically free similar to
nonabelian gauge theories. For $k\rightarrow \infty$ the dimensionless Newton constant
$g(k)\equiv k^2 G(k)$ approaches a non-Gaussian UV attractive fixed point $g^{\rm UV}_\ast$.
This means that $G(k)$ vanishes proportional to $1/k^2$ for $k\rightarrow\infty$.
The non-Gaussian fixed point of 4-dimensional quantum gravity is similar to the Weinberg fixed point
in $2+\epsilon$ dimensions \cite{wein}.

In the following we shall use the known results about the running of $G(k)$ and
$\Lambda(k)$ in order to ``renormalization group improve'' the Einstein equations
which govern the evolution of the Universe. They contain Newton's constant $G$ and the 
cosmological constant $\Lambda$. The improvement is done by substituting 
$G\rightarrow G(k)$, $\Lambda\rightarrow \Lambda(k)$, and by expressing $k$ 
in terms of the geometrically relevant IR cutoff. Considering only homogeneous and
isotropic cosmologies we shall argue that the correct identification of the 
cutoff is $k\propto 1/t$ where $t$ is the cosmological time. 

Similar RG improvements are standard tools in particle physics. A first gravitational
RG-improvement based upon the effective average action has been described in 
refs.\cite{bh1,bh2} where quantum effects in black hole spacetimes were studied.

In the present paper we shall set up a system of differential equations which consists
of the RG equations for $G$ and $\Lambda$, the improved Einstein equations, an 
additional consistency condition dictated by the Bianchi identities, and the equation
of state of the matter sector. This system determines the evolution of $G$, 
$\Lambda$, $a$, $\rho$ and $p$ as a function of the cosmological time $t$. We shall 
see that for $t \searrow 0$   all solutions to this system have a simple power law
structure. This attractor-type solution fixes $\rt=\rc$ without any finetuning. 
If the matter system is assumed to obey the equation of state of ordinary radiation, 
the scale factor expands linearly, $a(t)\propto t$, so that the RG-improved spacetime
has no particle horizon. For $t$ much larger than the Planck time the solutions
of the RG-improved system approach those of standard FRW cosmology. 

The remaining sections of this paper are organized as follows. In Section II we review
the essential properties of the effective average action for gravity and the 
solutions of its RG equation which we need in the present context. In Section III we describe 
the derivation of the RG improved Einstein equations and in Section IV we obtain 
solutions to it which are valid
for $t\rightarrow 0$ and $t\rightarrow \infty$, respectively. In section V we
investigate the physical properties of solutions which are valid during the entire Planck era.
In Section VI we discuss the generation of primordial density perturbations
and Section VII contains the Conclusions. 

In the main body of this paper we use a specific identification of the cutoff $k$ 
in terms of the cosmological time ($k\propto 1/t$). In appendix A we compare 
the results to those obtained with a different cutoff ($k\propto 1/a(t)$). 
In the main part of the paper we improve the basic {\it equations} for the cosmological
evolution. In appendix B we describe the alternative strategy of improving the 
{\it solutions} to the classical equations. 
\section{The effective average action for gravity}
In this section we review some properties of the effective average action 
$\Gamma_k[g_{\mu\nu}]$ and collect various results which we 
shall need in the present investigation. The average action for gravity has been constructed
in \cite{mr} using an approach which 
in earlier work \cite{ym,ym2,sven,ber} had already been tested for Yang-Mills theory. 

The definition of $\Gamma_k[g_{\mu\nu}]$ is based upon a modified gauge fixed path-integral of
$d$-dimensional Euclidean gravity in the background gauge. The crucial new ingredient is an 
IR cutoff which suppresses the contributions from long-wavelength metric fluctuations with momenta
smaller than $k$. In a second step, the functional $\Gamma_k$ defined by the modified 
path-integral is shown to satisfy an exact functional differential equation, the flow
equation, from which $\Gamma_k$, for all values of $k$, can be computed if it is 
known at some initial point $\widehat k$. In order to obtain an action $\Gamma_k[g]$ which is 
invariant under general coordinate transformations the standard background gauge formulation
has been employed. This leads to the complication that we actually have to RG-evolve
an action $\Gamma_k[g,\bar{g}]$ which depends on both  the ``ordinary'' metric $g_{\mu\nu}$ 
and on the background metric $\bar{g}_{\mu\nu}$. The standard action with one argument is recovered by setting 
$\bar{g} = g$, {\it i.e.} $\Gamma_k[g]\equiv\Gamma_k[g,{g}]$. The flow equation for 
$\Gamma_k[g,\bar{g}]$ reads \footnote{This is already a simplified form
of the flow equation appropriate for truncations which neglect the running 
of the ghost term. For its most general form see \cite{mr}.}
\ba\label{2.1}
&&k\partial_k \Gamma_k[g,\bar{g}]=\nonumber\\ 
&&{\frac{1}{2}}{\rm Tr}\Big [ \kappa^{-2}(\Gamma_k^{(2)}[g,\bar{g}]+{\cal R}^{\rm grav}_k[\bar{g}] )^{-1}
k\partial_k {\cal R}^{\rm grav}_k[\bar{g}]\Big ]
-{\rm Tr}\Big [( -{\cal M}[g,\bar{g}]+{\cal R}^{\rm gh}_k[\bar{g}])^{-1}k\partial_k
{\cal R}_k^{\rm gh}[\bar{g}]\Big ]
\ea
where $\Gamma_k^{(2)}$ stands for the Hessian of $\Gamma_k$ with respect to $g_{\mu\nu}$
and ${\cal M}$ is the Faddeev-Popov ghost operator. The operators ${\cal R}^{\rm grav}_k$
and ${\cal R}^{\rm gh}_k$ implement the IR cutoff in the graviton and the ghost sector.
They are defined in terms of a to some extent arbitrary smooth function 
${\cal R}_k(p^2)\propto k^2 {R}^{(0)}(p^2/k^2)$ by replacing the momentum square $p^2$
with the graviton and ghost kinetic operator, respectively. Inside loops, they suppress the
contribution of infrared modes with covariant momenta $p<k$. The function $R^{(0)}(z),\;$
$z\equiv p^2/k^2$, has to satisfy the conditions $R^{(0)}(0)=1$ and $R^{(0)}(z)\rightarrow 0$
for $z\rightarrow \infty$. For explicit computations the exponential cutoff
\be\label{2.2}
R^{(0)}(z)=z [{\rm exp}(z) -1]^{-1}
\ee
is particularly convenient. 

In order to find approximate but nonperturbative solutions to the flow equation the 
``Einstein-Hilbert truncation'' had been adopted in \cite{mr}. This means that the RG flow
in the space of all actions is projected onto the two-dimensional subspace spanned by 
$\int \sqrt{g}$ and $\int \sqrt{g} R$. This truncation of the ``theory space'' amounts to 
considering only actions of the form \footnote{In \cite{mr} the notation $G_k \equiv G(k)$
and $\bar{\lambda}_k \equiv \Lambda(k)$ had been used.} 
\be\label{2.3}
\Gamma_k[g,\bar{g}]=
(16\pi G(k))^{-1}\int d^dx \sqrt{g} \{ -R(g) + 2\Lambda(k)\} + \text{classical gauge fixing}
\ee
where $G(k)$ and $\Lambda(k)$ denote the running Newton constant and cosmological
constant, respectively. More general (and, therefore, more precise) truncations would include higher powers
of the curvature tensor as well as nonlocal terms \cite{nonloc}, for instance. 
By inserting (\ref{2.3}) into (\ref{2.1}) and performing the projection we obtain a coupled
system of equations for $G(k)$ and $\Lambda(k)$. It is most conveniently written down in terms
of the dimensionless Newton constant
\be\label{2.4}
g(k)  \equiv k^{d-2} \;G(k)
\ee
and the dimensionless cosmological constant 
\be\label{2.5}
\lambda(k) \equiv \Lambda(k) /k^2
\ee
One finds
\be\label{2.6}
k\partial_k \;g = \left[d-2+ \eta_N \right]g
\ee
and
\be\label{2.7}
\bay\dis
k\partial_k\;\lambda
&=&\dis
-(2-\eta_N)\lambda+\frac{1}{2}g (4\pi)^{1-d/2}\cdot
\\
&&\dis
\quad
\cdot\left[
2d(d+1) \Phi^1_{d/2}(-2\lambda)-8d \Phi^1_{d/2}(0)
-d(d+1)\eta_N\widetilde\Phi^1_{d/2}(-2\lambda)\right]
\eay
\ee
Here
\be\label{2.8}
\eta_N(g,\lambda)= \frac{ g B_1(\lambda)}{ 1-g B_2(\lambda)}
\ee
is the anomalous dimension of the operator $\sqrt{g} R$, and the functions $B_1(\lambda)$ and
$B_2(\lambda)$ are given by
\be\label{2.9}
\bay\dis
B_1(\lambda)
&\equiv&\dis
\frac{1}{3}(4\pi)^{1-d/2}
\Bigg[
 d(d+1) \Phi^1_{d/2-1}(-2\lambda)
-6d(d-1)\Phi^2_{d/2}(-2\lambda)
\\
&&\dis
\qquad\qquad
\quad
-4d\Phi^1_{d/2-1}(0)-24\Phi^2_{d/2}(0)\Bigg]
\\
B_2(\lambda)
&\equiv&\dis
-\frac{1}{6}(4\pi)^{1-d/2}
\left[
 d(d+1) \widetilde\Phi^1_{d/2-1}(-2\lambda)
-6d(d-1)\widetilde\Phi^2_{d/2}(-2\lambda)
\right]
\eay
\ee
with the threshold functions 
$(p=1,2,\cdots)$ 
\be\label{2.10}
\bay\dis
\Phi^p_n(w)
&=&\dis
\frac{1}{\Gamma(n)}\int_0^\infty dz\,
z^{n-1}
\frac{R^{(0)}(z)-z R^{(0)\,\prime}(z)}{[z+R^{(0)}(z)+w]^p}
\\
\widetilde\Phi^p_n(w)
&=&\dis
\frac{1}{\Gamma(n)}\int_0^\infty dz\,
z^{n-1}
\frac{R^{(0)}(z)}{[z+R^{(0)}(z)+w]^p}
\eay
\ee
These equations are valid for an arbitrary spacetime dimension $d$. In the following
we shall focus on the case $d=4$.

Clearly it is not possible to find solutions to the system (\ref{2.6}), 
(\ref{2.7}) in closed form; for a numerical determination of the phase diagram 
we refer to \cite{frank}.
However, for our purposes it will be sufficient to know
the behavior of the solutions in the limiting cases $k\rightarrow 0$ and 
$k\rightarrow \infty$. For small values of the cutoff the solutions are power
series in $k$. For the dimensionful quantities one obtains \cite{mr}
\ba\label{2.11}
&&G(k) = G_0\; [ 1-\omega \;G_0 k^2 + O(G_0^2 k^4)]\\[2mm]
&&\Lambda(k) = \Lambda_0 +\nu \;G_0 k^4 \;[ 1+O(G_0 k^2)]\label{2.12}
\ea
with the constants 
\ba\label{2.13}
&&\omega =\frac{1}{ 6\pi} \; [ 24 \; \Phi^2_2(0) -\Phi^1_1(0)]\\[2mm]
&&\nu = \frac{1}{ 4 \pi}\; \Phi^1_2(0)\label{2.14}
\ea
As it stands, (\ref{2.12}) for $\Lambda(k)$ is correct only if one either neglects the 
backreaction of the running $\Lambda$ via the $\Phi$-functions, or if one chooses 
$\Lambda_0=0$. For $\Lambda_0 > 0$ and with the backreaction due to the argument
of $\Phi^1_2(-2\Lambda/k^2)$ included, the RG trajectory runs into
a singularity and cannot be continued below a certain critical value of $k$.
This is probably due to the fact that the Einstein-Hilbert truncation is too simple to describe
the IR behavior of quantum gravity with a positive cosmological constant. Since
in this paper we are mostly interested in UV-physics we avoid this problem
by restricting ourselves to the case $\Lambda_0=0$.

The precise values of $\omega$ and $\nu$ depend on the choice of the cutoff
function $R^{(0)}$. For every admissible $R^{(0)}$ both constants are positive,
however. In (\ref{2.11}) and (\ref{2.12}) we wrote $G_0 \equiv G(k=0)$ and 
$\Lambda_0\equiv \Lambda(k=0)$ for the infrared values of $G$ and $\Lambda$.
At least within the Einstein-Hilbert truncation, $G(k)$ does not run any more
between scales where Newton's constant has been determined experimentally
(laboratory scale, scale of the Solar System, etc.) and the cosmological scale
where $k\approx 0$. Therefore we may identify $G_0$ with the experimentally observed
value of Newton's constant. We use $G_0$ in order to define the (conventional)
Planck mass $\mp$, Planck length $\lp$ and Planck time $\tp$:
\be\label{2.15}
\mp = G_0^{-1/2},\;\;\;\;\;\;\; \lp=\tp= G_0^{1/2}
\ee
The solutions (\ref{2.11}) and (\ref {2.12}) are expansions in the dimensionless
ratio $(k/\mp)^2$. Obviously the renormalization effects become strong
only if $k$ is about as large as $\mp$. We see that $G(k)$ decreases when we 
increase $k$ which is a first hint at the asymptotic freedom
of pure quantum gravity \cite{mr}.

In the following we shall say that $k$ is in the {\it perturbative regime}
if the approximations (\ref{2.11}) and (\ref{2.12}) are valid, {\it i.e.}
if $k\lesssim \mp$ so that the first order in the $(k/\mp)$-expansion
is sufficient to describe the running of $G$ and $\Lambda$.

Next let us look at the opposite limiting case when $k\gg \mp$. It turns out 
\cite{souma,bh2,frank,ol} that for $k\rightarrow \infty$ the physically relevant RG trajectories
in $(g,\lambda)$-space run into a UV-attractive fixed point  
$(g_\ast^{\rm UV},\lambda_\ast^{\rm UV})$. For the exponential cutoff (\ref{2.2}) the 
numerical analysis \cite{souma,frank,ol} of (\ref{2.6}), (\ref{2.7}) yields
the values $g_\ast^{\rm UV}\approx 0.27$ and $\lambda^{\rm UV}_\ast \approx 0.36$.
(If one neglects the running of $\lambda$ there is still a fixed point for $g$
at $g_\ast^{\rm UV}\approx 0.71$.) The existence of this fixed point implies that 
for $k\gg \mp$ the dimensionful quantities run according to 
\ba\label{2.16}
&&G(k) = \frac{g_\ast^{\rm UV} }{ k^2}\\[2mm]
&&\Lambda(k) = \lambda_\ast^{\rm UV} \; k^2\label{2.17}
\ea
We observe that for $k\rightarrow \infty$ Newton's constant, and hence the strength 
of the gravitational interaction, decreases very rapidly so that gravity is
``asymptotically free''. In fact, $G$ runs much faster than the gauge coupling 
constant in Yang-Mills theory which depends on $k$ only logarithmically. An
asymptotic running of the form (\ref{2.16}) had been conjectured by Polyakov \cite{pol}.
A similar power-like running of $G$ has already been known to occur in 
$(2+\epsilon)$-dimensional gravity \cite{wein,mr}. In fact, the fixed point
$(g_\ast^{\rm UV},\lambda_\ast^{\rm UV})$ is the 4-dimensional counterpart of 
Weinberg's fixed point in $2+\epsilon$ dimensions \cite{souma}. If the existence of the
fixed point can be confirmed by more general truncations this means that Einstein
gravity in 4 dimensions is ``asymptotically safe'' and as
well-behaved and predictive as a perturbatively renormalizable theory \cite{wein}.
In this paper we assume that the general picture provided by the Einstein-Hilbert
truncation is at least qualitatively correct, and that the UV fixed point
does indeed exist. In fact, recent investigations including an $R^2$-term in the 
truncation \cite{ol} indicate that in the UV the Einstein-Hilbert truncation indeed
captures all the essential physics.

We shall say that $k$ is in the {\it fixed point regime} if $k\gg \mp$
so that the asymptotic solutions (\ref{2.16}), (\ref{2.17}) apply. 

For intermediate values of $k$ the RG equations can be solved numerically only. 
However, if one neglects the influence of $\Lambda$ on the running of $G$ 
(and omits a tiny correction coming from $B_2(0)$ ) one obtains
the following simple formula which is valid for all $k$ \cite{bh2}:
\be\label{2.18}
G(k) = {G_0 \over 1+\omega \; G_0 k^2}
\ee
For $k$ small we recover (\ref{2.11}), and for $k^2\gg G_0^{-1}$ the fixed 
point behavior sets in, $G(k)\approx 1/\omega k^2$, 
so that $G(k)$ becomes independent of its IR value $G_0$. 

Up to now we discussed pure gravity without matter fields. But of course any matter 
field leads to an additional renormalization of $G$ and $\Lambda$ 
\cite{percacci,odintsov}. In \cite{percacci} the average action approach has been 
generalized and an arbitrary number of free scalars, spinors, vector fields, and
Rarita-Schwinger fields has been added. (See also \cite{alfio,dal}.)
Depending on the nature and on the number of the matter fields gravity either continues
to be antiscreening and asymptotically free, or the quantum effects of the matter 
fields overwhelm those of the metric and destroy asymptotic freedom. (The same 
happens in QCD with too many quark flavours.) In this paper we assume that
the matter system is such that the resulting RG-flow for $G$ and $\Lambda$ is 
qualitatively the same as in pure gravity. In particular, we assume that there is
a non-Gaussian fixed point which is UV attractive for $g$ and $\lambda$,
but we allow the numerical values of
$g_\ast^{\rm UV}$ and $\lambda_\ast^{\rm UV}$ to differ from their pure gravity values.
In fact, none of our conclusions will depend on the values
$g_\ast^{\rm UV}$, $\lambda_\ast^{\rm UV}$, $\omega$, and $\nu$ provided all those parameters
are strictly positive.
 
In the sequel we shall write $g_\ast$ and $\lambda_\ast$ for $g_\ast^{\rm UV}$ and 
$\lambda_\ast^{\rm UV}$, respectively.
\section{The RG improved Einstein equations}
We consider homogeneous, isotropic cosmologies described by Robertson-Walker metrics
of the form
\be\label{3.1}
ds^2=-dt^2+a(t)^2\Big [{dr^2\over 1-Kr^2}+r^2(d\theta^2+\sin^2\theta \;d\phi^2)\Big ]
\ee
For $K=0$ the $3$-spaces of constant cosmological time $t$ are flat, 
and for $K=+1$ and $-1$ they are spheres and pseudospheres, respectively. In standard FRW
cosmology the dynamics of the scale factor $a(t)$ is determined by
Einstein's equations
\be\label{3.2}
R_{\mu\nu}-{1\over 2} g_{\mu\nu}R = -\Lambda g_{\mu\nu} + 8\pi G\;T_{\mu\nu}
\ee
where $G$ and $\Lambda$ are constant parameters. In order to take the 
leading quantum corrections into account we are now ``improving'' (\ref{3.2})
by replacing $G$ and $\Lambda$ with the scale dependent quantities $G(k)$ and $\Lambda(k)$.

In general it is a difficult task to identify the actual physical cutoff mechanism which, in 
a concrete situation, stops the running in the infrared. Typically this involves
expressing $k$ in terms of all scales which are relevant to the problem under consideration,
such as the momenta of particles, field strengths, or the curvature of the spacetime, for
instance. In the case at hand the situation simplifies because the conditions of 
homogeneity and isotropy imply that $k$ can be a function of the cosmological
time only: $k=k(t)$. Provided we know how $k$ depends on $t$ we can turn the solutions of
the RG equation, $G(k)$ and $\Lambda(k)$, into functions of time:
\be\label{3.3}
G(t)\equiv G(k=k(t)), \;\;\;\;\; \Lambda(t) \equiv \Lambda(k=k(t))
\ee 
There are two plausible scales which could determine the identification of $k$ in terms
of $t$. The first one is $k\propto 1/t$. In fact, the temporal proper distance of some
point $P(t,r,\theta,\phi)$ to the big bang (which still will be present in the 
improved spacetime) is directly given by $t$ itself. If we want to construct
an effective field theory $\Gamma_{k}$ which is valid near $P$ we may not integrate 
out quantum fluctuations with momenta smaller than $1/t$ because by the time
the age of the Universe is $t$, fluctuations with frequencies smaller than $1/t$ 
cannot have played any role yet. By this argument we are indeed led to the identification
\be\label{3.4}
k(t)={\xi\over t}
\ee
where $\xi$ is a positive constant. (Note that $t$ and $a$ have 
mass dimension $-1$, while $r,\theta,\phi,K$ and $\xi$ are dimensionless.) As it stands,
(\ref{3.4}) refers to the $t,r,\theta,\phi$-coordinate system, but it has an invariant
meaning. At any point $P$ we set
\be\label{3.4b1}
k(P)={\xi\over d(P)}
\ee
where $d(P)\equiv \int _{{\cal C}(P)}\sqrt{ds^2}$ is the proper length of the curve 
${\cal C}(P)$ as given by the metric (\ref{3.1}). With respect to the 
$t,r,\theta,\phi$-system, ${\cal C}(P)$ is defined by $\lambda\mapsto (\lambda,r,\theta,\phi)$
with $\lambda\in [0,t]$ where $(t,r,\theta,\phi)$ are the coordinates of $P$.
Both the metric and the curve can be re-expressed in a generic coordinate system $x^\mu$,
so that the cutoff is actually a scalar function $k(x^\mu)$.

Another momentum scale which appears natural at first sight is 
\be\label{3.5}
k(t)= {\xi\over a(t)},
\ee
but in particular for the most important case of $K=0$ it is not obvious why the RG
flow should be stopped at this point. In fact, it will turn out that  
for the perturbative regime the improved system of equations has no consistent solution
if one uses (\ref{3.5}). On the other hand, for the fixed point regime of a 
radiation-dominated Universe (\ref{3.4})  and (\ref{3.5}) lead to exactly the same
answers so that our predictions are particularly robust in this case. A third scale 
one might invoke is the Hubble parameter 
\be\label{3.6}
H(t)={\dot{a}(t)\over a(t)}
\ee
However, in the present context only power laws $a\propto t^\alpha$ are of interest.
For them $H$ is proportional to $1/t $ and does not define an independent scale.

While we believe that the leading effects are correctly described by the 
$1/t$-cutoff, the more subtle sub-leading effects most probably require more
complicated cutoffs which, apart from an {\it explicit} time dependence, also have
an {\it implicit} time dependence via $a(t)$ and its derivatives:
\be\label{3.6b1}
k = k(t, a(t), \dot{a}(t), \ddot{a}(t),\cdots )
\ee
In this paper we discard those sub-leading effects. From now on we assume that 
$k\propto 1/t$ is indeed the correct first order approximation and we shall use
(\ref{3.4}) in the main body of the paper. For comparison we also investigate 
the consequences of the $1/a$-cutoff (\ref{3.5}) in Appendix A.

Upon inserting (\ref{3.4}) into (\ref{2.11}) and (\ref{2.12}) we obtain for the time
dependent Newton constant and cosmological constant in the perturbative
regime
\ba\label{3.7}
&&G(t) = G_0 \; \Big [1-\widetilde{\omega} \Big ( {\tp\over t}\Big )^2+O\Big({\tp^4\over t^4}\Big) \Big ]\\[2mm]
&&\Lambda(t) = \Lambda_0 + \widetilde{\nu} \;\mp^2 
\Big ({\tp\over t}\Big )^4\Big [1+O\Big ( {\tp^2\over t^2}\Big )\Big ]\label{3.8}
\ea
with the positive constants
\be \label{3.9}
\widetilde{\omega}\equiv\omega \;\xi^2, \;\;\;\; \widetilde{\nu}\equiv \nu \;\xi ^4
\ee
In the fixed point regime we get from (\ref{2.16}), (\ref{2.17}) 
\ba\label{3.10}
&&G(t)= \widetilde{g}_\ast \;t^2\\[2mm]
&&\Lambda(t) = \widetilde{\lambda}_\ast \; t^{-2}\label{3.11}
\ea
with 
\be\label{3.12}
\widetilde{g}_\ast\equiv g_\ast \; \xi^{-2},\;\;\;\;\; \widetilde{\lambda}_\ast\equiv
\lambda_\ast \; \xi^2
\ee
In order to find the functions $G(t)$ and $\Lambda(t)$ which interpolate between the
behavior (\ref{3.7}), (\ref{3.8}) and (\ref{3.10}), (\ref{3.11}) one must solve the 
RG equation numerically.

The next issue is the energy momentum tensor $T_{\mu\nu}$ to be used on the RHS of
the improved Einstein equations. Because of the imposed homogeneity and isotropy 
it can always be transformed to the form
\be\label{3.13}
{T_{\mu}}^\nu= {\rm diag} (-\rho, p, p, p)
\ee
where the density $\rho$ and the pressure $p$ depend on $t$ only. As in standard cosmology we assume
that the energy-momentum tensor is covariantly conserved\footnote{See for instance 
ref.\cite{mrcw} for a class of cosmologies with a time dependent $\Lambda$ where $T_{\mu\nu}$ as defined here
is not conserved.},
\be\label{3.14}
D_\nu {T_\mu}^\nu= 0
\ee
so that for the Robertson-Walker metric
\be\label{3.15}
\dot{\rho}+3\;{\dot{a}\over a}\;(\rho+p)=0.
\ee
The physical picture behind $T_{\mu\nu}$ is not necessarily that of a perfect classical
fluid as in the familiar FRW case. We rather interpret it as the functional derivative
of some {\it effective} action $\Gamma^{\rm M}[g_{\mu\nu}]$ for the matter system in the
background of the metric $g_{\mu\nu}$. For the equation of state relating $p$ to $\rho$ we shall use the linear
ansatz 
\be\label{3.16}
p(t)= w\; \rho(t)
\ee
where $w$ is an arbitrary constant. It includes the case of a perfect fluid consisting of 
classical dust $(w=0)$ or radiation $(w=1/3)$, but we emphasize that $\Gamma^{\rm M}$ is by no
means restricted to describe classical matter. In particular, $w$ may be different from
its classical value.

Let us return to the Einstein equation (\ref{3.2}) now. By virtue of Bianchi's identity its LHS
is covariantly conserved, so for consistency the RHS must be conserved too:
\be\label{3.17}
D_\nu [ -\Lambda {g_\mu}^\nu + 8\pi G \; {T_\mu}^\nu] = 0
\ee
Because $\Lambda$ and $G$ depend on $t$, this equation is not automatically satisfied
if $T_{\mu\nu}$ is conserved. Instead we obtain the following consistency condition
which relates the time dependencies of $\Lambda$, $G$ and $\rho$:
\be\label{3.18}
\dot{\Lambda} +8\pi\rho\; \dot{G} = 0
\ee
Sometimes it is convenient to rewrite (\ref{3.18}) in the form
\be\label{3.19}
{d\over dt}(\Lambda + 8\pi G\;\rho) = 8\pi G \;\dot{\rho}
\ee

When we insert the Robertson-Walker metric (\ref{3.1}) into Einstein's equation (\ref{3.2}) 
we obtain two independent equations:
\be\label{3.20}
\Big ( {\dot{a}\over a} \Big )^2 + {K\over a^2} = 
{1\over 3}\;\Lambda+{8\pi\over 3}G\;\rho
\ee
from the $00$-component, and
\be\label{3.21}
2\;{\ddot{a}\over a}+\Big ({\dot{a}\over a}\Big )^2+{K\over a^2}= 
\Lambda-8\pi G\;\rho
\ee
from the {\it ii}-components. As in the classical case, these two field equations
are consistent only if $T_{\mu\nu}$ is conserved. After multiplying 
(\ref{3.20}) by $a^2$, taking its time derivative, and combining it with 
(\ref{3.21}) one obtains the conservation law (\ref{3.15}) as an integrability 
condition for the improved Einstein equations. In this calculation essential use is made of 
the new consistency condition (\ref{3.19}). We see that its role is completely analogous
to the conservation equation for $T_{\mu\nu}$: both of them constrain the sources to 
which gravity can be coupled consistently. Thus only 2 of the 3 equations (\ref{3.15}), 
(\ref{3.20}) and (\ref{3.21}) are independent; in the following we shall use the 
conservation law (\ref{3.15}) and the improved Friedmann equation (\ref{3.20}) 
as independent equations.

To summarize: we would like to write down a set of (differential) equations
which determine $a,\rho,p,G$ and $\Lambda$ as a function of time. This set 
includes Friedmann's equation, the conservation law for $T_{\mu\nu}$, the
equation of state, the new consistency condition, and the RG equations for $G$ and
$\Lambda$. More precisely, we shall always assume that the RG equations are already solved so that
we can simply replace the constant $k$ by $k(t)$ in the solution. Eliminating the pressure
by virtue of the equation of state, this system of equations reads
\begin{subequations}
\ba
&&\Big ({\dot{a}\over a}\Big )^2+{K\over a^2}= 
{1\over 3}\Lambda+{8\pi\over 3} G\rho\label{3.22a}\\[2mm]
&&\dot{\rho}+3(1+w){\dot{a}\over a}\rho= 0\label{3.22b}\\[2mm]
&&\dot{\Lambda}+8\pi\rho \;\dot{G} = 0\label{3.22c}\\[2mm]
&&G(t)= G(k(t)),\;\;\; \Lambda(t) = \Lambda(k(t))\label{3.22d}
\ea
\end{subequations}
These are 5 equations for the 4 functions $ a(t), \rho(t), G(t)$ and $\Lambda(t)$.
(Of course we could immediately insert (\ref{3.22d}) into the first 3 equations.
Then (3.24a,b,c) are 3 equations for the 2 unknowns $a$ 
and $\rho$. For the time being we shall not adopt this point of view.)

The system (3.24a,b,c) without the last equations coming from the renormalization
group has already been studied in the literature \cite{sys1,sys2}.
It consists of only 3 equations for 4 unknowns and is underdetermined therefore.
As a way out, the authors made an ad hoc assumption about one of the functions,
typically $G(t)$, and checked if there are interesting cosmologies consistent
with, but not uniquely determined by (3.24a,b,c).

In our case with Eq.(\ref{3.22d}) included we seem to be in the opposite situation
because the 5 equations might overdetermine the 4 unknowns and no consistent solution
might exist. In order to see that this is not the case actually we must return to the RG
equation from which (\ref{3.22d}) is derived. The flow equation contains the function
$R^{(0)}$ which is completely arbitrary up to the two conditions 
$R^{(0)}(0)=1$ and $R^{(0)}(z\rightarrow \infty)=0$. This function describes the details
of the cutoff mechanism, {\it i.e.} how quickly the modes with different momenta $p$ get 
suppressed when $p$ approaches $k$. Only if one uses the flow equation in order to compute 
quantities which are ``universal'' in the sense of statistical mechanics
the answers are independent of the shape of $R^{(0)}$. In general
$\Gamma_k$, for intermediate values of $k$, does depend on $R^{(0)}$. 
(Only the limit $k\rightarrow 0$ is $R^{(0)}$-independent because the cutoff
drops out.) Therefore the RG trajectory $k\mapsto (G(k),\Lambda(k))$ is also
$R^{(0)}$-dependent. This is obvious from the Eqs.(\ref{2.13},\ref{2.14}),
for instance: the coefficients $\omega$ and $\nu$ depend on $R^{(0)}$ via the
$\Phi$-integrals. This means that if we want to give a physical meaning 
to $G(k)$ and $\Lambda(k)$ at intermediate values of $k$, 
the function $R^{(0)}$ should be chosen in such a way that it models the
actual {\it physical} cutoff mechanism as accurately as possible. 

Similarly, 
also the identification of the scale $k$ in terms of the actual physical
parameters of the system depends on the system under consideration. In our case
we have $k=\xi/t$ with an unknown constant $\xi$. If we change $R^{(0)}$ also the
optimal value for $\xi$ changes. Typically combinations of parameters in the 
RG equation $(\omega, \nu, \cdots)$ and in the cutoff identification $(\xi)$
such as $\widetilde{\omega}=\omega\xi^2$, for instance, are much less $R^{(0)}$-dependent,
${\it i.e.}$ more ``physical'', than those parameters separately.
(For the RG improved Newton potential it can be checked that the 
$R^{(0)}$-dependences of $\omega$ and an analogously defined $\xi^2$ mutually cancel,
and that $\widetilde\omega$ is a physical, {\it i.e.} observable quantity
\cite{bh2}.) However, even measurable combinations similar to $\widetilde\omega$
cannot be calculated by RG techniques alone. 

In this situation it is a virtue of the system (3.24) rather than 
a disadvantage that it is seemingly overdetermined because in this manner
it places restrictions also on $R^{(0)}$ and on the cutoff identification. In fact, we
may regard it as a system of 5 integro-differential equations for the 5 functions
$a, \rho, G, \Lambda$ and $R^{(0)}$. In the next section we shall solve this system
in the perturbative and in the fixed point regime, and we shall see that solutions
exist only if certain relations among the parameters $\widetilde\omega$,
$\widetilde{g}_\ast$, etc. are satisfied. They are implicit conditions on
$R^{(0)}$ and/or $\xi$. This shows that the system (3.24) is quite powerful
in the sense that it also teaches us something about how to consistently model
the IR cutoff for the concrete system ``expanding Universe''. 

This enhanced degree of predictability is also one of the reasons why 
we are RG improving {\it equations} rather than {\it solutions}.
Improving solutions means that we take some fixed solution $a(t)$, $\rho(t)$ 
of standard cosmology which depends parametrically on the constants $G$ and $\Lambda$
and then substitute $G\rightarrow G(t)$, $\Lambda\rightarrow \Lambda(t)$.
In general this simple approach is reliable only if the improved solution is close to the
classical one. (See \cite{bh2} for a detailed discussion in the context of 
black holes.) The main advantage of improving the underlying equations is that 
their solutions may well be quite different from the classical ones without necessarily
lying in a domain where the entire approach has become unreliable.
In appendix B we describe the improvement of the classical FRW solutions. Where they 
are valid, the results are consistent with the approach of improving equations. They are 
less predictive, however, in particular because they do not reproduce the relations
among $\widetilde\omega$, $\widetilde{g}_\ast$, etc. mentioned above.

It is important to understand how many constants of integration occur in the process of 
solving the system (3.24). Let us pick some $R^{(0)}$ and a function $k=k(t)$ 
with an explicit $t$-dependence only. Then $G(k)$ and $\Lambda(k)$ can be obtained by solving
2 coupled RG equations which are of first order and lead to 2 constants of integration
therefore. We choose them to be the $k=0$-values $G_0$ and $\Lambda_0$. As a consequence, 
the functions $G(t)$ and $\Lambda(t)$ in (\ref{3.22d}) depend parametrically
on $G_0$ and $\Lambda_0$, {\it i.e.} on the RG trajectory selected. In a first step we 
may insert (\ref{3.22d}) into (\ref{3.22c}) and obtain the energy density as
\be\label{3.23}
\rho(t) = -{1\over 8 \pi} \; {\dot{\Lambda}\over \dot{G}}
\ee
The time dependence of $\rho$ is completely determined once $\Lambda(t)$ and $G(t)$ are fixed,
and no new constant of integration arises. In a second step we insert $\rho$ of (\ref{3.23}) 
into (\ref{3.22b}) and solve the resulting differential equation for $a(t)$. Eq.(\ref{3.22b}) 
is easily integrated:
\be\label{3.24}
\rho(t) \; [a(t)] ^{3+3w} = {\cal M}/8\pi = {\rm const.}
\ee
Here we encounter a further constant of integration, ${\cal M}$. Its  mass dimension is 
$1-3w$. For a radiation dominated Universe ${\cal M}$ is dimensionless, while it has the dimension
of a mass in the matter dominated case. Combining (\ref{3.23}) and (\ref{3.24}) we obtain
the scale factor 
\be\label{3.25}
a(t) = \Big [ \; -{{\cal M} \; \dot{G}\over \dot{\Lambda}}\;\Big ] ^ {1/(3+3w)}
\ee
Already at this point all 4 functions $G$, $\Lambda$, $\rho$ and $a$ are completely determined.
They depend on 3 constants of integration: $G_0$, $\Lambda_0$ and ${\cal M}$. The last and
crucial step is to insert the solution we found into (\ref{3.22a}) and check if this equation is
satisfied too. In general it will be satisfied only for appropriately chosen cutoff functions 
$R^{(0)}$ and $k(t)$, and for special values of the constants of integration and of the parameter $w$.

We note that also the Hubble parameter has a simple representation directly in terms of $G$ and
$\Lambda$:
\be\label{3.26}
H={\dot{a}\over a} = {1\over {3+3w}}\; \Big ( {\ddot{G}\over \dot{G}}-{\ddot{\Lambda}\over\dot{\Lambda}}\Big )
\ee

It is clear that the system (3.24) can be solved in this simple manner only in the special
case when $k(t)$ has no implicit time dependence via $a(t)$. For a generic $k=k(t, a(t), \cdots)$ the
situation is much more involved, see for instance Appendix A for the ansatz $k=\xi/a$.

Before closing this section let us introduce a few convenient definitions. We define the 
vacuum energy density $\rl$, the total energy density $\rt$ 
and the critical energy density $\rc$ according to  
\ba\label{3.27}
&&\rl(t) \equiv {\Lambda (t) \over 8\pi G(t)}\\[2mm]
&&\rt(t) \equiv \rho + \rl\label{3.28}\\[2mm]
&&\rc(t) \equiv {3\over 8 \pi G(t) }\Big ( {\dot{a}\over a}\Big )^2\label{3.29}
\ea
The definitions (\ref{3.27}) and (\ref{3.29}) are the same as usual except that $G(t)$ and 
$\Lambda(t)$ appear in place of $G_0$ and $\Lambda_0$. This means in particular that 
for very late times when the running Newton constant assumes its IR value $G_0$, 
the quantity $\rc$ is exactly the standard critical density of classical FRW cosmology.
It is also customary to introduce
\ba\label{3.30}
&&\Omega_{\rm M} \equiv {\rho\over \rc},\;\;\;\;\;\;\; \oa \equiv {\rl\over\rc}\\[2mm]
&&\ot \equiv {\Omega_{\rm M} + \oa}={\rt\over\rc}\label{3.31}
\ea
so that we may rewrite Friedmann's equation (\ref{3.22a})
either as
\be\label{3.32}
{\dot{a}^2+K\over a^2}= {8\pi\over 3}\;G\;\rt
\ee
or as
\be\label{3.33}
K=\dot{a}^2\;\Big [ {\rt\over \rc}-1\Big ] = \dot{a}^2 \;\Big [ \ot -1\Big ]
\ee

As a trivial consequence of its definition, the critical density satisfies
\be\label{3.34}
\rc (t) \; G(t) \; H(t) ^{-2} = {3\over 8\pi}
\ee
By Eq.(\ref{3.33}), an expanding Universe with $K=0$ has
\be\label{3.35}
\rt(t) = \rc(t) \;\;\;\;\;\;\;\;\;\;\;\;\;\;\;\;\;\;\;\;\;\;\;\; (K=0)
\ee
at any time. In this case 
\be\label{3.36}
\rt (t) \; G(t) \; H(t)^{-2} = {3\over 8\pi } \;\;\;\;\;\;\;\;\;\;\;\;\; (K=0)
\ee
Sometimes the flatness problem is rephrased as the cosmological ``coincidence puzzle'':
Why does the product of the observed matter density of the Universe, the square of its age $t$, 
and of Newton's constant give rise to a number of order unity, 
\be\label{3.37}
(\rho \; G \; t^2)_{\rm today} = O(1)  \;\;\;\;\;\;\;\;\;?
\ee
It is clear that (\ref{3.37}) is essentially the same statement as (\ref{3.36}) 
if $\rl$ is negligible or at most of the same order of magnitude as $\rho$, 
and if the age of the Universe is of the order of $H(t)^{-1}$. The ``coincidence''
(\ref{3.37}) has also been regarded as a manifestation of Mach's principle \cite{mach}.

\section{perturbative and fixed point solutions}

In this section we solve the system of equations (3.24) using the approximated 
RG equations which are valid in the perturbative and in the fixed point regime,
respectively.

\subsection{The perturbative regime}
The perturbative approximation is valid for $k\ll \mp$, {\it i.e.} for 
$t\gg \tp$. The corresponding solutions to the RG equations are given by 
(\ref{3.7}), (\ref{3.8}) from where we obtain 
\ba\label{4.1}
&&\dot{G}(t)={2\;\widetilde{\omega}\;G_0^2\over t^3} \; \Big \{1
+O\Big ({\tp^2 \over t^2}\Big) \Big \}\\[2mm]
&&\dot{\Lambda}(t) = - {4\; \widetilde{\nu}\;G_0\over t^5}\; \Big \{ 1+ 
O\Big ({\tp^2\over t^2}\Big )\Big \} \label{4.2}
\ea
Hence Eq.(\ref{3.23}) for the energy density and Eq.(\ref{3.25}) for the scale factor lead to
\be
\rho(t) = {1\over 4\pi}\; \Big ( {\widetilde{\nu}\over \widetilde{\omega} }\Big )\;
{1\over G_0 t^2}\; \Big \{ 1+O\Big ({\tp^2\over t^2}\Big ) \Big \} \label{4.3}
\ee 
and 
\be\label{4.4}
a(t) = \Big [ {1\over 2} \; \Big ( {\widetilde{\omega}\over \widetilde{\nu}}\Big )
\; {\cal M} \; G_0 \Big ]^{1/(3+3w)}\; t^{2/(3+3w)}\; \Big \{ 
1+O\Big ({\tp^2 \over t^2} \Big ) \Big \}, 
\ee
respectively. Now we must insert (\ref{4.3}) and (\ref{4.4}) along with $G(t)$ 
and $\Lambda(t)$ from (\ref{3.7}) and (\ref{3.8}) into the Friedmann equation
(\ref{3.22a}) in order to check whether the above solutions are consistent. 
Omitting sub-leading terms, consistency requires that
\be\label{4.5}
\Big( {2\over 3+3w}\Big )^2{1\over t^2}+K
\Big [{1\over 2}\Big ({ \widetilde{\omega}\over \widetilde{\nu}}\Big )
{\cal M}G_0 \Big ]^{-2/(3+3w)} {1\over t^{4/(3+3w)}}=
{\Lambda_0\over 3}+ \Big ( {8\pi G_0 \over 3}\Big )
{1\over 4\pi}\Big ( {\widetilde{\nu}\over \widetilde{\omega}}\Big )
{1\over G_0 t^2}+\cdots
\ee
Note that on the RHS of (\ref{4.5}) it is sufficient to set $G=G_0+\cdots$
and $\Lambda = \Lambda_0+\cdots$ because the (known) corrections to these
approximations have the same time dependence as the (unknown) second order
corrections on the LHS. In order to analyze Eq.(\ref{4.5}) we must distinguish
the cases $K=0$ and $K=\pm 1$. 

{\bf a) The case $K=0$.}

\noindent
In the case $K=0$, Eq.(\ref{4.5}) is fulfilled provided that the consistency conditions
\be\label{4.6}
\Lambda_0 =0 \;\;\;\;\; \text{and} \;\;\;\;\;{\widetilde{\omega}\over \widetilde{\nu}} =
{3\over 2} (1+w)^2
\ee
are satisfied. The condition $\Lambda_0=0$ does not come as a surprise because the formula
(\ref{2.12}) for $\Lambda(k)$ from which we started is accurate for $k\rightarrow 0$
only if $\Lambda_0=0$. Recalling that $\widetilde{\nu}/\widetilde{\omega}
=(\nu/\omega)\xi^2$ we see that the second condition puts a constraint on the cutoff
$R^{(0)}$ which affects $\omega$ and $\nu$, as well as the function $k=k(t)$,
{\it i.e.} $\xi$ in our case. We use this condition in order to express 
$\xi$ in terms of $\omega$ and $\nu$ which are not subject to any further condition then:
\be\label{4.7}
\xi^2 = {2 \omega \over 3\nu(1+w)^2}
\ee
Thus, upon inserting (\ref{4.7}) into (\ref{3.7}) and (\ref{3.8}), the time
dependence of Newton's constant and of the cosmological constant is completely
determined now. Moreover, using (\ref{4.6}) for the ratio 
$\widetilde{\omega}/\widetilde{\nu}$ in Eq.(\ref{4.3}) and Eq.(\ref{4.4}) we see that 
$\rho(t)$ and $a(t)$ are actually completely independent of $\omega$ and 
$\nu$. As a consequence, the consistent solution we found is given by the following four
equations:
\begin{subequations}
\ba\label{4.8a}
&&a(t) = \Big [ {3\over 4}(1+w)^2{\cal M}G_0\Big ]^{1/(3+3w)}\;t^{2/(3+3w)}
\;\Big \{1+O\Big ({\tp^2\over t^2}\Big ) \Big \}\\[2mm]
&&\rho(t) = {1\over 6\pi (1+w)^2 G_0 t^2}+O\Big ({1\over t^4}\Big )\label{4.8b}\\[2mm]
&&G(t) = G_0 \Big [1-{2\omega^2\over 3\nu(1+w)^2}
\Big ({\tp\over t}\Big ) ^2 +O\Big ({\tp^4\over t^4}\Big )\Big ]\label{4.8c}\\[2mm]
&&\Lambda(t) = {4\omega^2\mp^2\over 9\nu (1+w)^4}\Big ({\tp\over t}\Big )^4+
O\Big ({\tp^4\over t^6}\Big )\label{4.8d}
\ea
\end{subequations}
We observe that the leading terms of the above expressions for $a(t)$ and $\rho(t)$ 
coincide {\it exactly} with the corresponding solutions of the classical FRW equations.
(See Eqs.(\ref{B.3}) and (\ref{B.4}) in Appendix B.)
This coincidence is quite remarkable because in our approach, by Eqs.(\ref{3.23}) and 
(\ref{3.25}), $a$ and $\rho$ arise from the {\it time dependent}, 
{\it i.e.} higher order terms in $G(t)$ and $\Lambda(t)$, which clearly
have no counterpart in the classical situation. 

The vacuum energy density and the critical energy density for the cosmology (4.8) are 
\be\label{4.9}
\rl = 0+O\Big ( {1\over t^4} \Big ), \;\;\; \rc = \rho +O\Big ( {1\over t^4}\Big )
\ee            
so that, in leading order, $\rt=\rho=\rc$, or 
\be\label{4.10}
\Omega_{\rm M} =1, \; \; \; \; \Omega_{\Lambda}=0, \;\;\;\; \Omega_{\rm tot}=1 
\ee

{\bf b) The case $K=\pm 1$.}
 
\noindent
Equation (\ref{4.5}) has a chance of being consistent only if all terms can be given
a time dependence proportional to $1/t^2$. If $K\not= 0$ this is possible only 
for an ``exotic'' equation of state with $ w = -1/3$. Indeed the consistency conditions
implied by (\ref{4.5}) are
\be\label{4.11}
\Lambda_0=0, \;\;\;\; w = -{1\over 3}, \;\;\;\;
{\widetilde{\omega}\over\widetilde{\nu}}={2\over 3}-{2K\over {\cal M} G_0}
\ee
Again we use the last condition in order to eliminate $\xi$:
\be\label{4.12}
\xi^2 = {\omega\over\nu} \; \Big ( {2\over 3} - {2K\over {\cal M} G_0} \Big)^{-1}
\ee
Note that in the present case $\xi$ depends also on the constants of integration, 
${\cal M}$ and $G_0$. Proceeding as above we find the following consistent 
solution for $w=-{1/ 3}$:
\begin{subequations}
\ba\label{4.13a}
&&a(t)=\Big [ {1\over 3} {\cal M}G_0 -K\Big ]^{1\over 2} \; t \;
\Big \{ 1+O\Big ( {\tp^2\over t^2} \Big ) \Big \}\\[2mm]
&&\rho(t) = {{\cal M}\over 8 \pi}\Big [ {1\over 3} {\cal M}G_0 -K\Big ]^{-1}\;
{1\over t^2}\; \Big \{ 1+O\Big ( {\tp^2\over t^2} \Big ) \Big \}\label{4.13b}\\[2mm]
&&G(t) = G_0 \Big [1-{\omega^2\over \nu}
\Big ( {2\over 3}-{2K\over  {\cal M} G_0} \Big )^{-1}
\; \Big ( {\tp\over t}\Big )^2 + O\Big ({\tp^4\over t^4} \Big ) \Big ]\label{4.13c}\\[2mm]
&&\Lambda(t) = {\omega^2 \over \nu } \; \mp^2 \;
\Big ( {2\over 3}-{2K\over  {\cal M} G_0} \Big )^{-2}
\; \Big ( {\tp\over t} \Big )^4 + O\Big ({\tp^4\over t^6} \Big )\label{4.13d}
\ea
\end{subequations}
The leading terms in (\ref{4.13a}) and (\ref{4.13b}) coincide with the corresponding 
classical FRW solutions for $w = -1/3$. The cosmology (4.13) gives rise to
\be\label{4.14}
\rl(t) = 0 + O\Big ({1\over t^4}\Big) \;\;\;\; 
\rc (t) = {3\over 8\pi G_0 t^2}+O\Big ( {1\over t^4} \Big )
\ee
so that, in leading order, $\Omega_{\Lambda}=0$ and
$\Omega_{\rm M} = \Omega_{\rm tot}$ with 
\be\label{4.15}
\Omega_{\rm tot} = {{\cal M} G_0 \over {\cal M}G_0 -3K}
\; \Big \{ 1+O\Big ( {\tp^2\over t^2} \Big ) \Big \}
\ee
As expected, $\Omega_{\rm tot}$ depends on the constants of integration
in the case $K= \pm 1$. 
\subsection{The fixed point regime}
The fixed point approximation is valid when $k\gg \mp$ or $t\ll \tp$. In this regime the 
time dependence of $G$ and $\Lambda$ is given by Eqs.(\ref{3.10}) and (\ref{3.11}), 
respectively. From (\ref{3.23}) and (\ref{3.25}) we obtain 
\ba\label{4.16}
&&a(t) = \Big ( {\widetilde{g}_\ast{\cal M}\over \widetilde{\lambda}_\ast} \Big )^{1/(3+3w)}
\; t^{4/(3+3w)}\\[2mm]
&&\rho (t) = {\widetilde{\lambda}_\ast\over 8\pi \widetilde{g}_\ast }
\; {1\over t^4}\label{4.17}
\ea
The next step is to check the consistency of (\ref{3.22a}). Inserting $G$, $\Lambda$
and the above expressions for $a$ and $\rho$ we have 
\be\label{4.18}
\Big ( {4\over 3+3w} \Big )^2 \; {1\over t^2}+
K\Big [ {\widetilde{g}_\ast{\cal M}\over \widetilde{\lambda}_\ast}\Big ]^{-2/(3+3w)}
{1\over t^{8/(3+3w)}}= {2\widetilde{\lambda}_\ast\over 3 t^2}
\ee
We shall discuss this equation for $K=0$ and $K=\pm 1$ separately.

{\bf a) The case $K=0$. }
 
\noindent
For $K=0$, Eq.(\ref{4.18}) implies only a single consistency
condition:
\be\label{4.19}
\widetilde{\lambda}_\ast = {8\over 3 (1+w)^2}
\ee
If we use this condition in order to eliminate $\widetilde{\lambda}_\ast$
in all equations we are led to 
\begin{subequations}
\ba\label{4.20a}
&&a(t) = \Big [ {3\over 8}(1+w)^2\; \widetilde{g}_\ast 
\; {\cal M} \Big ]^{1/(3+3w)} \; t^{4/(3+3w)} \\[2mm]
&&\rho(t) = {1\over 3 \pi (1+w)^2 \;\widetilde{g}_\ast} 
\; {1\over t^4}\label{4.20b}\\[2mm]
&&G(t) = \widetilde{g}_\ast \; t^2 \label{4.20c}\\[2mm]
&&\Lambda (t) = {8\over 3 (1+w)^2} \; {1\over t^2}\label{4.20d}
\ea
\end{subequations}
This family of solutions, one for each value of $\widetilde{g}_\ast$
and $w$, was found in ref.\cite{sys2} already. In this work, the RG equations
(\ref{3.22d}) had not been used. Since the system (\ref{3.22a}), 
(\ref{3.22b}), (\ref{3.22c}) is underdetermined, the time dependence for
$G(t)$, Eq.(\ref{4.20c}) above, had been postulated on an ad hoc basis in order
to obtain a unique solution. In this manner the analogue of $\widetilde{g}_\ast$
appears as a free parameter while $\widetilde{\lambda}_\ast$ is fixed. In our case 
it is more natural to use the consistency condition (\ref{4.19}) in order to express
$\xi$ in terms of $\lambda_\ast$ which is given by the renormalization group.
Because $\tla\equiv\lambda_\ast\xi^2$ we have then
\be\label{4.21}
\xi^2={8\over 3(1+w)^2\; \lambda_\ast}
\ee
When expressed in terms of the fixed point values, the solutions read
\begin{subequations}
\ba\label{4.22a}
&&a(t) = \Big [ \Big ({3\over 8}\Big )^2 (1+w)^4 \; g_\ast \lambda_\ast \; {\cal M} \Big ]^{1/(3+3w)}\; t^{4/(3+3w)}\\[2mm] 
&&\rho(t) = {8\over 9\pi (1+w)^4 \; g_\ast \lambda_\ast}\; {1\over t^4}\label{4.22b}\\[2mm]
&&G(t) = {3\over 8}(1+w)^2 \; g_\ast \lambda_\ast \; t^2\label{4.22c}\\[2mm]
&&\Lambda(t) = {8\over 3(1+w)^2}\; {1\over t^2}\label{4.22d}
\ea
\end{subequations}
Since $g_\ast$, $\lambda_\ast$ and $w$ are given by the renormalization group and the equation
of state, respectively, (4.22) represents a one-parameter family of solutions
parametrized by the constant ${\cal M}$. The solutions (4.22) reflect the renormalization
group flow in the vicinity of the UV attractive fixed point where the RG trajectories have ``forgotten''
their IR values $G_0$ and $\Lambda_0$. Because of this universality, these solutions are independent
of the constants of integration $G_0$ and $\Lambda_0$. This means that (4.22) is an 
attractor solution for  $t\searrow 0$ in the sense that {\it every} consistent solution to
(3.24), characterized by arbitrary constants of integration ($G_0$, $\Lambda_0$, 
${\cal M}$), looks like (4.22) in the limit $t \searrow 0$. Actually  the 
${\cal M}$-dependence of the solutions (4.22) is quite trivial: $\rho$, $G$ and $\Lambda$
are ${\cal M}$-independent, while $a(t)$ responds to a change of ${\cal M}$ by a simple 
constant rescaling.

It is very remarkable and a nontrivial confirmation of our approach that after the elimination
of $\xi$ the RG data enter the attractor solution only via the product $g_\ast\lambda_\ast$.
This product is {\it universal} (scheme-independent) 
in the sense that it does not depend on the function ${R}^{(0)}$
\cite{ol}. Hence (4.22) is free from any numerical ambiguities.

For the cosmologies (4.22) we find that $\rho_\Lambda (t) =\rho (t)$
and $\rho_{\rm crit}(t) = 2\rho(t)$ so that 
\be\label{4.23}
\rho=\rl={1\over 2}\; \rc, \;\;\;\;\; \rt = \rc
\ee
or
\be\label{4.24}
\Omega_{\rm M} = \Omega_{\Lambda} = {1\over 2}, \;\;\;\;\; \Omega_{\rm tot} = 1
\ee
We also read off the Hubble parameter
\be\label{4.25}
H={4\over 3+3w} \; {1\over t}
\ee
and observe that 
\be\label{4.26}
\rho(t)\; G(t)\; t^2 ={1\over 3\pi (1+w)^2}
\ee
is a time-independent fixed number which depends only on the equation of state.

The solutions (4.22) exists for every equation of state of the type considered, 
{\it i.e.} for every value of the parameter $w$. Since at least immediately after the 
Planck era during which (4.22) is valid the Universe is radiation dominated, a
particularly plausible choice is $w=1/3$. In the case of a ``radiation dominated 
Planck era'' with $w=1/3$ we have
\begin{subequations}
\ba
&&a(t) = \Big [ {4\over 9} \; g_\ast \lambda_\ast \; {\cal M} \Big ]^{1/4} \; t \label{4.27a}\\[2mm]
&&\rho(t)= {9\over 32\pi \;g_\ast \lambda_\ast}\; {1\over t^4}\label{4.27b}\\[2mm]
&&G(t) = {2\over 3} \; g_\ast \lambda_\ast \; t^2 \label{4.27c}\\[2mm]
&&\Lambda(t) = {3\over 2} \; {1\over t^2}\label{4.27d}
\ea
\end{subequations}
The most interesting property of this solution is that it is perfectly {\it scale free}.
Because ${\cal M}$ is dimensionless for $w=1/3$ and because $G_0$ and $\Lambda_0$ do not
occur due to the fixed point behavior, the only dimensionful quantity available is the 
cosmological time $t$ itself. As a consequence, the various exponents of $t$ appearing in 
(4.27) are completely fixed by the canonical mass dimensions of $a,\rho,G$ and $\Lambda$, which are 
$-1$,$+4$,$-2$, and $+2$, respectively. In particular, the linear expansion law $a\propto t$
is a direct consequence of this type of scale invariance. Since $w=1/3$ corresponds to a traceless 
energy momentum tensor, this solution is realized if $\Gamma^{\rm M}$ is the effective action of
a quantum conformal field theory, for instance.

{\bf b) The case $K= \pm 1$.}
 
\noindent
In this case Eq.(\ref{4.18}) can be made consistent only for a specific choice of the 
equation of state, namely for $w=+1/3$.
Eq.(\ref{4.18}) is satisfied if 
\be\label{4.28}
w=+{1\over 3} \;\;\;\;\;\text{and}\;\;\;\;\;
1+K\Big ({\widetilde{\lambda}_\ast \over \widetilde{g}_\ast {\cal M}} \Big )^{1/2}={2\over 3}\;
\widetilde{\lambda}_\ast
\ee
We use the second consistency condition in order to eliminate $\xi$ in
favour of $g_\ast$, $\lambda_\ast$ and ${\cal M}$:
\be\label{4.29}
\xi^2 = \Big ( {g_\ast {\cal M}\over \lambda_\ast }\Big )^{1\over 2}
\Big [{2\over 3} \sqrt{g_\ast \lambda_\ast {\cal M}}-K\Big ]^{-1}
\ee
This leads to the following solutions for $w=+1/3$:
\begin{subequations}
\ba
&&a(t)=\Big [ {2\over 3}\sqrt{g_\ast\lambda_\ast {\cal M}}-K\Big ]^{1/2}\;t\label{4.30a}\\[2mm]
&&\rho(t) = {{\cal M}\over 8\pi}\; \Big [ {2\over 3} \sqrt{g_\ast \lambda_\ast {\cal M}}-K\Big ]^{-2}\; {1\over t^4}
\label{4.30b}\\[2mm]
&&G(t) = \Big ( {g_\ast \lambda_\ast \over {\cal M} } \Big )^{1/2}
\; \Big [ {2\over 3} \sqrt{g_\ast \lambda_\ast {\cal M}}-K\Big ]\; t^2\label{4.30c}\\[2mm]
&&\Lambda(t) = \sqrt{g_\ast \lambda_\ast {\cal M}}\; 
\Big [ {2\over 3} \sqrt{g_\ast \lambda_\ast {\cal M}}-K\Big ]^{-1} \; {1\over t^2}\label{4.30d}
\ea
\end{subequations}
This family of solutions, again parametrized by a dimensionless constant ${\cal M}$, 
is scale free as well. All solutions have the property that their vacuum energy density equals the 
matter density:
\be\label{4.31}
\rl (t) = \rho (t) = {1\over 2} \; \rt (t)
\ee
Furthermore, their critical density reads 
\be\label{4.32}
\rc (t) = {3\over 8\pi} \; \Big ( {{\cal M}\over g_\ast \lambda_\ast} \Big )^{1/2}
\; \Big [ {2\over 3} \sqrt{g_\ast \lambda_\ast {\cal M}}-K\Big ]^{-1}\; {1\over t^4}
\ee
from which one obtains
\be\label{4.33}
\Omega_{\rm M} = \Omega_\Lambda = {1\over 2} \Omega_{\rm tot} = 
{1\over 3} \sqrt{g_\ast \lambda_\ast {\cal M}}\; \Big [ {2\over 3} \sqrt{g_\ast \lambda_\ast {\cal M}}-K\Big ]^{-1}
\ee
If $K=+1$, solutions of the form (4.30) exist only if ${\cal M}$ is such that 
$\sqrt{g_\ast\lambda_\ast {\cal M}}> 3/2$. It is also important to note that for 
$K=\pm1$ the quantity
\be\label{4.34}
\rho (t) \; G(t) \; t ^2 = {1\over 8\pi}\sqrt{g_\ast \lambda_\ast {\cal M}} \;
\Big [ {2\over 3} \sqrt{g_\ast \lambda_\ast {\cal M}}-K\Big ]^{-1}
\ee
is not a universal number but depends on ${\cal M}$. 
\section{Complete solutions for the Planck Era}
\subsection{Early vs. late stages of the Planck era}
In the previous section we found solutions to the RG improved system of cosmological 
evolution equations which are valid for $t\searrow 0$ and for $t \gtrsim \tp$, respectively.
In particular, it turned out that the improved cosmologies, too,  start from a ``big bang'', 
{\it i.e.} there exists a time (conveniently chosen as $t=0$) at which the scale factor vanishes.
We also saw that there is a certain transition time $t_{\rm class}$ such that for 
$t> t_{\rm class}$ quantum gravitational effects become negligible so that the evolution of the 
Universe is correctly described by the classical FRW models. The time $t_{\rm class}$ is of the 
order of a few Planck times, $t_{\rm class} \gtrsim \tp$. We shall refer to the epoch between 
$t=0$ and $t=t_{\rm class}$ as the {\it Planck era}. At the beginning of the Planck era, immediately
after the big bang, we are in the fixed point regime of the RG equations, while the end of the 
Planck era and its transition to classical cosmology corresponds to the perturbative 
regime. 

We were able to find analytic solutions to the improved equations only for the very early and the very 
late part of the Planck era. Let us now discuss how those solutions can be fitted together to obtain
complete solutions which are valid during the entire Planck era. 

For a spatially flat geometry, $K=0$, and for every value of $w$, there exist exact solutions of 
(3.24) both in the fixed point and in the perturbative regime, 
see Eqs.(4.22) and (4.8), respectively. We expect that those two limiting 
solutions possess a continuous interpolation which satisfies the equations (3.24) 
for all $t\in (0, t_{\rm class})$. Generically this interpolating solution should exist, because we have 
considerable freedom in adjusting the functions ${R}^{(0)}$ and $k(t, a(t),\cdots)$
without changing their qualitative features. We shall refer to this solution
$\{a(t), \rho(t) , G(t), \Lambda(t) \}$ , $t\in (0, t_{\rm class})$,  as the 
{\it complete $K=0$ solution}. Actually this is a whole family of solutions labeled by the constants
of integration $(G_0, \Lambda_0, {\cal M})$. (Within the present approximation, only solutions with 
$\Lambda_0 = 0$ were found.) 

It is the main assumption of this paper that the RG improved system (3.24)
and its complete $K=0$ solution are valid throughout the Planck era, {\it i.e.} even immediately
after the big bang. The reason why we think that our approximations are valid even for 
$t\searrow 0$ is the asymptotic freedom we found for quantum gravity. It entails gravity 
in the very early Universe being weakly coupled. In fact, the coupling constant, 
{\it i.e.} Newton's constant vanishes very rapidly as we approach 
the initial singularity: $G\propto t^2$. For $k\rightarrow \infty $
the RG flow in $(g,\lambda)$-space is dominated by a fixed point which is UV attractive
for both $g$ and $\lambda$. By the RG improvement,
this fixed point translates into the attractor solution (4.22) for $a,\rho,G$ and $\Lambda$. 
In the vicinity of the attractor, all solutions have the same universal behavior. 

The $w$-value of the perturbative regime must coincide with the one of the following 
classical era, most plausibly $w=1/3$. In principle it is conceivable that the interpolation from the 
fixed point to the perturbative regime involves an adiabatic change of $w$. 

For the spatially curved geometries with $K=+1$ or $-1$ we found a solution in the fixed point
regime only if $w=+{1/3}$, and a solution in the perturbative regime only for $w=-1/3$. Hence, at least 
within the present approximation, there exists no consistent interpolating solution for 
$t\in (0, t_{\rm class})$ with a constant $w$. 

As for the interpretation of this result, we must be 
very careful. Clearly it would be premature to conclude that the RG approach predicts 
$K=0$ as the only possibility. In particular the nonexistence of perturbative solutions with $w\not = -1/3$
is quite likely to be an artifact of our approximations. We mentioned already that the simple 
perturbative form of $\Lambda(t)$, Eq.(\ref{2.12}), is correct only if one
either neglects the backreaction of the running $\Lambda$ contained in the $\Phi$-functions or if 
one specializes to $\Lambda_0 =0 $. In general the situation is similar to QCD \cite{gluco}
where, thanks to asymptotic freedom, simple truncations are sufficient for large values of $k$, but at small 
$k$ they necessarily become very complicated because they have to describe all sorts of nonperturbative
effects. On the basis of this analogy we expect that also in quantum gravity it is much more difficult 
to correctly describe the IR behavior. It is intriguing that in our approach this problem is 
particularly pressing if $\Lambda_0\not= 0 $. In fact, it has been suggested \cite{tsamis}
that there are strong renormalization effects in the IR which might solve the cosmological constant
problem \cite{cos} in a dynamical way. 

It is less obvious why for $K= \pm 1$ there seem to be no solutions with $w\not = +1/3$ in the fixed point
regime. It would be tempting to speculate that this reflects a property of the exact theory in which case
the slightest deviation from the classical value $w=+{1/3}$ would lead to the prediction that $K=0$. 
\subsection{``Naturalness'' of the solutions}
Let us now make more precise in which sense the existence of the complete ($K=0$) RG improved
solution removes the flatness problem. We emphasize that the reason is {\it not}
that we found no solutions for $K=\pm 1$ and that $\rt =\rc$ is automatic if $K=0$. In fact, for the 
sake of the argument, let us suppose that there is some better approximation 
(an exact treatment) such that there are complete solutions also for $K=\pm 1$
and perhaps also for $K=0$, $\Lambda_0\not =0$. Then, both the classical and the RG improved theory describe 
cosmologies with all 3 types 
of spatial geometries: flat $(K=0)$, spherical $(K=+1)$, and pseudospherical $(K=-1)$. Let us select one out
of these 3 options, $K=0$ say, and let us compare what the two theories have to say about the
evolution of the Universe. 

{\it Classical FRW cosmology} has a limited domain of applicability. It is valid only for $t\ge t_{\rm i}$ 
where $t_{\rm i} \gtrsim t_{\rm class}$ is some initial time at which one must specify initial conditions for the 
classical differential equations. They include the initial density $\rho ( t_{\rm i} )$ and the Hubble parameter
$H(t_{\rm i})$ from which one can deduce the initial critical density $\rho_{\rm crit}(t_{\rm i}) \equiv
3 H(t_{\rm i})^2/8\pi G_0$. Since we opted for $K=0$, the classical differential equations tell us
that there is a solution only if the initial conditions are such that $\rho ( t_{\rm i})= 
\rho_{\rm crit}(t_{\rm i})$.
Thus, in order to be in the $K=0$-sector, an infinite finetuning of the initial data is necessary, 
and this is what is referred to as the flatness problem.

Because gravity is weakly coupled for $t\searrow 0$, {\it RG improved cosmology} has the ambition of being
valid for all $t>0$, {\it i.e.} already directly after the big bang. At $t=0$ the spacetime is singular, and 
there is no such thing as a $t=t_{\rm i}$-hypersurface at which initial data are to be imposed. 
There is a family of complete consistent $K=0$ cosmologies labeled by the parameters 
$(G_0, \Lambda_0, {\cal M})$. For any value of the parameters, $\rt (t) = \rc (t)$ is automatically
satisfied for all $t>0$. For $t\searrow 0$ all solutions approach an essentially 
universal attractor solution which is independent of  $(G_0, \Lambda_0, {\cal M})$
except for an overall ${\cal M}$-dependence of $a$. It is precisely this attractor
which makes it not only unnecessary but even impossible to specify initial conditions in a standard way.
Thus, by the time when the classical solution 
emerges from the quantum solution, {\it the condition $\rho_{\rm tot}=
\rho_{\rm crit}$ is imposed automatically}. 

To summarize: At present the RG improvement provides no strong theoretical arguments against 
$K=+1$ or $-1$. However, if one selects the $K=0$-option ``by hand'', no naturalness problem occurs.



\begin{figure}
\hbox to\hsize{\hss\epsfxsize=18cm\epsfbox{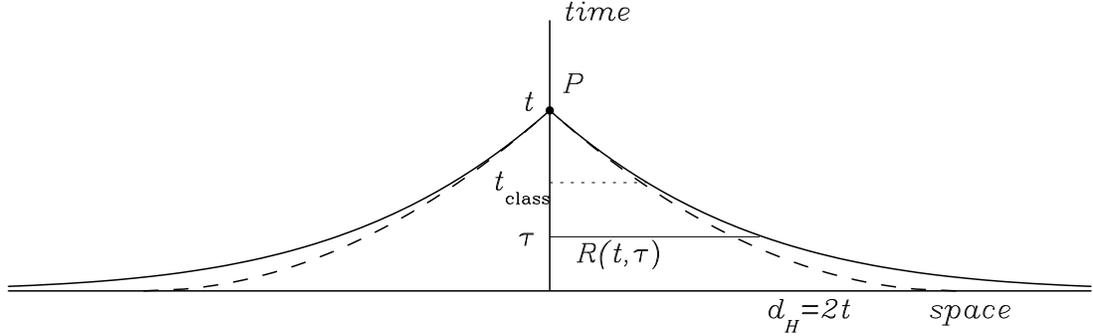}\hss}
\caption{ Graphical representation of the proper distance $R(t,\tau)$ as a function of $\tau$
for fixed $t$. Only light signals emitted from points below the solid line can reach the
spacetime point $P$. The dashed line shows $R_{\rm class}(t,\tau)$ which gives rise to a horizon
at $d_H = 2t$. The deviation of $R$ from $R_{\rm class}$ becomes appreciable only for
$\tau< t_{\rm class}$.
\label{fig.1}}
\end{figure}

\begin{figure}
\hbox to\hsize{\hss\epsfxsize=18cm\epsfbox{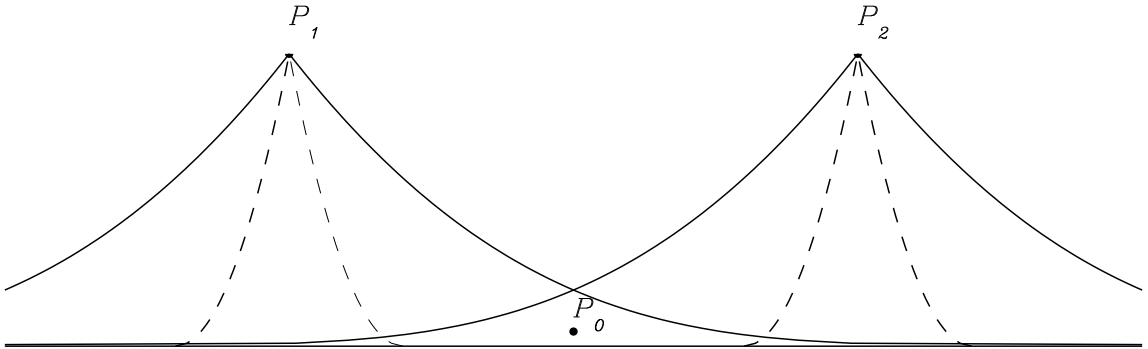}\hss}
\caption{ While the points $P_1$ and $P_2$ are causally disconnected classically, the
quantum gravity induced broadening of the backward light cones allows for
events $P_0$ in the Planck era
which can causally influence both $P_1$ and $ P_2 $ . 
\label{fig.2}}
\end{figure}

\subsection{Particle horizons}
Let us consider an observer in a Robertson-Walker spacetime who, at cosmological time $t$, receives a light 
signal which was emitted by some distant galaxy at time $\tau<t$. Then, at time $t$, the proper distance 
between this galaxy and the observer is given by \cite{weinbook}
\be\label{5.1}
R(t,\tau) = a(t) \int^t_\tau {dt'\over a(t')}
\ee
In a spacetime with a singularity at time zero, the most distant galaxies from which the observer
can receive a light signal at time $t$ have the proper distance $R(t,0)\equiv d_H(t)$.
If this distance is finite, {\it i.e.} if the integral (\ref{5.1}) converges for $\tau \rightarrow 0$, 
\be\label{5.2}
d_H(t) = a(t) \int^t_0 {dt'\over a(t')} 
\ee 
we say that the spacetime has a particle horizon at the distance $d_H$. Hence it is the 
$t\searrow 0$ behavior of the scale factor which decides about the presence or absence 
of a particle horizon. For instance, if 
\be\label{5.3}
a(t) \propto t^\alpha\;\;\;\;\; (\alpha >0)
\ee
there is a horizon at $d_H(t) = t/(1-\alpha)$ for $\alpha \in (0,1)$ but there is no horizon if 
$\alpha\geq 1$. 

For $t\ll 1/\sqrt{\Lambda}$ all classical FRW solutions are power laws
of the type (\ref{5.3}) with the exponent
\be\label{5.4}
\alpha_{\rm class} = {2\over 3+3w}
\ee
(See Appendix B.) 
If we take these solutions at face value even for $t\searrow 0$, there appears to be a horizon
in both of the physically relevant cases of the radiation and the matter dominated Universe 
with $w=1/3$ and $w=0$, respectively. However, since the classical equations become
invalid for $t\searrow 0$ there is no compelling reason 
why these horizons actually should exist in Nature. 

In the RG improved cosmology for $K=0$ the early part of the Planck era is governed 
by the attractor solution (4.22) with 
\be\label{5.5}
a(t) \; \propto \; t^{4/( 3+3w)}
\ee
Since we believe that this attractor provides a valid description for 
$t\searrow 0$, even very close to the 
big bang, we may use (\ref{5.5}) in order to check for the existence of horizons. We observe that 
{\it the RG improved spacetime has no particle horizon provided $w\leq 1/3$.}
\footnote{Within the phenomenological applications \cite{sys1,sys2} of the 
system (3.24a,b,c) this has already been pointed out earlier \cite{sys2}.}

During the following  discussion we assume that the matter system is such that $w\leq 1/3$
so that there is indeed no horizon. However, as we shall see now, this fact by itself is not yet a solution
to the horizon problem. For the sake of simplicity we consider a ``radiation dominated Planck era''
with $w = 1/3$ followed by a classical radiation dominated era, again with 
$w=1/3$. In this case we have a linear expansion at early times and the 
familiar square-root expansion
at late times:


\be
a(t) \  \propto  \   \left\{
\begin{array}{ll} 
t           & \mbox{for $t \ll t_{\rm class} $} \\
t^{1/2}     & \mbox{for $t \gg t_{\rm class} $}
\\
\end{array}
\right.
\ee

In order to visualize the causal properties of this Robertson-Walker spacetime we 
consider a simple toy model which interpolates smoothly between $a\propto t$ for 
$t\ll t_{\rm class}$ and $a\propto t^{1/2}$ for $t\gg t_{\rm class}$:
\be\label{5.7}
a(t)= {A t \over 1+\sqrt{t/t_{\rm class}}}
\ee 
Here $A$ is an arbitrary positive constant. It is easy to calculate the proper distance 
(\ref{5.1}) for (\ref{5.7}):
\be\label{5.8}
R(t,\tau) = {t\over 1 +\sqrt{t/t_{\rm class}}}\Big [ {\rm ln}\Big ({t\over \tau}\Big )
+2\sqrt{t\over t_{\rm class}} - 2\sqrt{\tau\over t_{\rm class} } \Big ]
\ee
As expected, this distance diverges for $\tau \rightarrow 0$ and $t$ fixed. In Fig.\ref{fig.1} 
it is represented graphically as a kind of gravitationally distorted backward light
cone of the point $P$. It is compared to its classical counterpart
\be\label{5.9}
R_{\rm class}(t,\tau) = 2\sqrt{t}(\sqrt{t}-\sqrt{\tau})
\ee
which results from $a\propto t^{1/2}$ and gives rise to the familiar horizon at $d_H=2t$.

In Fig.\ref{fig.2} we show two spacetime points $P_1$ and $P_2$ at the same cosmological
time $t$. In classical cosmology those two points would be causally disconnected because their
``light cones'' given by $R_{\rm class}$ do not intersect. However, in the RG improved spacetime, 
the light cones become infinitely broad for $t\searrow 0$. 
This means that events which take place at sufficiently early times
$\tau$ can causally influence both $P_1$ and $P_2$. Because of this quantum gravity induced 
broadening of the backward light cones, the light cones 
of all events $P_i$ at a given time $t$ overlap for some small enough $\tau$. Since 
this broadening sets in only for $\tau \lesssim t_{\rm class} = O(\tp)$ 
we see that only events $P_0$ in the Planck 
era can causally influence {\it all} points $P$ on the hypersurface at time $t$.

Let us imagine, for instance, the two points $P_1$ and $P_2$ being located in opposite directions in the sky. 
Two microwave antennas pointing in these directions receive radiation that has been emitted at the time 
$t_{\rm r}$ 
of the Hydrogen recombination when the the cosmological plasma had just become optically thin to radiation, 
about $10^5$ years after the big bang. In the standard FRW-spacetime 
the number of horizon distances separating the two sources in 
opposite directions is given by 
\be\label{hd}
N = {2 R(t_0,t_{\rm r})\over d_H(t_{\rm r})} = 
\lim_{\tau\rightarrow 0} { 2 R(t_0,t_{\rm r})\over R(t_{\rm r},t_{\rm e})+ R(t_{\rm e},\tau)}
\ee
where $t_0$ denotes the present time, and  $t_{\rm e} \gtrsim t_{\rm class}$ is in the equivalence era, 
when matter and radiation were in local thermodynamic
equilibrium. However, since  both $R(t_0,t_{\rm r})$ and $R(t_{\rm r},t_{\rm e})$ 
are finite, it is clear that, in the quantum gravity improved spacetime,
eventually  $N<1$ for sufficiently small $\tau < \tp$.

In view of the above discussion we propose that the isotropy of the cosmic microwave background radiation
on large angular scales is a consequence of 
the quantum gravity effects in the Planck era which remove the particle horizon and hence allow for 
causal mechanisms giving rise to approximately the same temperature everywhere on the last scattering surface. 
The important point of this discussion is that since the broadening of the 
light cones becomes significant only for $t<\tp$, it is necessary that  those causal mechanisms 
are operative during the Planck era already.
In the following section we outline a scenario for the generation of primordial density
fluctuations where this is actually the case.
\section{Density fluctuations}
It is a fascinating idea that the structure formation in the Universe started out from primordial 
density fluctuations $\delta\rho({\mathbf{x}})$ which were triggered by quantum mechanical 
fluctuations. As the Universe expanded, those density fluctuations got amplified 
and magnified, and finally gave rise to the large-scale structures which we observe today. 
This idea has been worked out in the framework of inflationary cosmology. Here instead we 
consider the possibility that the primordial density fluctuations were generated already
during the Planck era as the aftermath of the big bang. This hypothesis allows us to invoke the broadening 
of the light cones for $t<t_{\rm class}$ which we found above in order to explain the high degree
of isotropy of the fluctuations at later times. 

In our approach the most natural assumption about the quantum origin of $\delta \rho$ is that, 
before $t\approx t_{\rm class}$, the quantum fluctuations of the metric itself generated the 
primordial density fluctuations by some decoherence mechanism. As we shall argue now, this 
assumption naturally leads to a scale free (Harrison-Zeldovich) fluctuation spectrum.

We need to know the two-point correlation function \cite{padbook}
\be\label{6.1}
\xi({\mathbf{x}}) = \langle \delta ({\mathbf{x}}+{\mathbf{y}}) \delta({\mathbf{y}}) 
\rangle
\ee
of the density contrast $\delta({\mathbf{x}})\equiv \delta\rho({\mathbf{x}})/\langle \rho\rangle_t$
at some {\it fixed} time $t\lesssim t_{\rm class}$ close to the end of the Planck era when 
the spectrum is ``handed over'' from the quantum gravity to the classical regime. We define
the power spectrum by
\be\label{6.2}
|\delta_k|^2\; \equiv V\int d^3 x \; \xi({\mathbf{x}}) 
\; e^{-i{\mathbf{k}}\cdot{\mathbf{x}}} 
\ee
and we say that the fluctuation spectrum has the spectral index $n$ if $|\delta_k|^2$ has the form of a 
power law $|\delta_k|^2\propto |{\mathbf{k}}|^n$. (V denotes the normalization volume.) What is the 
prediction for $|\delta_k|^2$ if our above hypothesis is correct?

In \cite{ol} it was shown that, on a flat background, the effective graviton propagator
for the fixed point regime is proportional to $\widetilde{{\cal G}}(p)\propto 1/p^4$ which amounts to
${\cal G}(x,y)\propto {\rm ln}(x-y)^2$ in position space. This form of the propagator is valid for
$p^2\gg\mp^2$ or $(x-y)^2\ll\lp^2$, respectively. The logarithmic two-point function may be understood
as a limiting case of the familiar ``critical'' propagator ${\cal G}(x,y)\propto 1/|x-y|^{d-2+\eta}$
for $d=4$ and the anomalous dimension $\eta\equiv \eta_N(g_\ast,\lambda_\ast) = -2$ which characterizes the
UV fixed point \cite{ol}. Let us look at the curvature fluctuation 
$\delta{\mathbf R}\propto \partial\partial h$
caused by a fluctuation $h_{\mu\nu}(x)$ of the metric. (We use a symbolic notation where ${\mathbf R}$ 
stands for the
curvature scalar or for any component of the Riemann or Einstein tensor.) Because 
$\langle h_{\mu\nu}(x) h_{\lambda\tau} (y) \rangle \propto {\rm ln}(x-y)^2$, the curvature correlation function
is $\langle \delta {\mathbf R}(x) \delta {\mathbf R}(y) \rangle \propto 1/(x-y)^4$, rather than 
$\propto 1/(x-y)^6$ as implied by the tree level propagator. Therefore the leading short distance singularity in 
a curved spacetime is given by $\langle \delta {\mathbf R}(x) \delta{\mathbf R}(y) \rangle \propto 
1/d(x,y)^4$ where $d(x,y)$ is the geodesic distance of $x$ and $y$. This formula is applicable when the spacetime
curvature is small compared to $1/d(x,y)^2$. 

Now we consider the background of a Robertson-Walker spacetime
and we put $x$ and $y$ on the same time slice. Hence $d(x,y)=a(t)|{\mathbf{x}}-{\mathbf{y}}|$
where ${\mathbf{x}}$ and ${\mathbf{y}}$ are the comoving Cartesian coordinates of $x$ and $y$,
respectively. This leads to the important result
\be\label{6.3}
\langle \delta {\mathbf R}({\mathbf{x}},t)\;\delta{\mathbf R}({\mathbf{y}},t)\rangle \propto 
{1\over |{\mathbf{x}}-{\mathbf{y}}|^4}
\ee
The constant of proportionality implicit in (\ref{6.3}) is time dependent but for the derivation of the 
spectrum this is unimportant. 

In the scenario where the primordial density fluctuations are generated by quantum fluctuations one assumes 
\cite{padbook} that the classical statistical expectation value (\ref{6.1}) is proportional to a 
quantum mechanical expectation value $\langle \Psi | \hat{\phi}({\mathbf{x}}+{\mathbf{y}})
\hat{\phi}({\mathbf{y}}) | {\Psi} \rangle$ where $\hat{\phi}$ is the operator whose
fluctuations are supposed to become classical. In the case at hand where we assume that 
$\delta\rho$ originates from the fluctuations of the spacetime geometry
itself the natural choice for $\hat{\phi}$ is $\hat{\phi}\propto {\mathbf R}$,
{\it i.e.} a to some extent arbitrary linear combination of curvature components. In fact, already 
classically the Einstein equation (\ref{3.2}) implies 
$8\pi G \; \delta\rho = - \delta {G_0}^0$ where ${G_\mu}^\nu$ is the Einstein tensor
\footnote{See ref.\cite{anton} for a similar discussion.}. As a consequence, the 
two-point function of $\hat{\phi}$ is proportional to the $\delta {\mathbf R}$ correlator (\ref{6.3}). Therefore
the correlation function of $\delta{\rho}$ behaves as 
\be\label{6.4}
\xi({\mathbf{x}})\propto {1\over |{\mathbf{x}}|^4}
\ee
provided the physical distance $a(t)|{\mathbf{x}}|$ is smaller than $\lp$. The power spectrum
of the modes with physical momenta $|{\mathbf{k}}|/a(t) \lesssim \mp$  (at fixed time
$t\lesssim t_{\rm class}$) is given by the 3-dimensional Fourier transform of
(\ref{6.4}):
\be\label{6.5}
|\delta_k|^2\propto |{\mathbf{k}}|
\ee
This is precisely the Harrison-Zeldovich scale invariant spectrum with the spectral index $n=1$.

We can thus imagine that ``sub-Hubble scale'' modes evolve according to the standard theory of
cosmological perturbations starting with a scale-invariant spectrum immediately
after the quantum gravity epoch, $t\gtrsim \tp$. 
A more complete treatment would include also the contribution from ``super-Hubble scale'' modes
in a gauge-invariant framework, but this is beyond the scope of the present
paper.

\section{Conclusion}
In this paper we studied homogeneous, isotropic cosmologies in the Planck era before the classical Einstein
equations become valid. We performed a RG improvement of the cosmological evolution equations by taking into 
account the running of $G$ and $\Lambda$ as it follows from the flow equation of the effective average action.
For a spatially flat geometry we found solutions to the improved equations which are mathematically consistent
even for $t\searrow 0$, {\it i.e}, immediately after the initial singularity of the the Universe. We believe that 
calculations can be done reliably in this regime because gravity becomes
asymptotically free at high momentum scales so that Newton's constant is very small close to the big bang:
$G\propto t^2$. The situation is comparable to QCD where physics at small length scales is simple
but becomes increasingly complex as one probes larger distance scales.
For $t\searrow 0$ the cosmological evolution is described by an attractor-type solution in 
$(a,\rho,G,\Lambda)$-space which is a direct manifestation of the UV fixed point of the RG flow 
in $(g,\lambda)$-space.

For a radiation dominated Planck era the attractor is perfectly scale free, the only dimensionful parameter being the 
cosmological time $t$. The RG improved solutions are ``natural'' in the sense that no finetuning is required, 
and for a broad class of equations of state ($w\leq 1/3$) they are free from particle horizons. 
Thus they offer an intriguing possibility 
for overcoming the flatness and the horizon problem of standard cosmology.
We also found a natural mechanism for generating a scale free spectrum of primordial
density fluctuations.

It is important to keep in mind which assumptions went into our derivation. They enter at different stages of 
the construction:
\begin{enumerate}

\item[{\it (i)}] We assume that for $k\rightarrow \infty$ the RG flow in $(g,\lambda)$-space is governed by a UV attractive
fixed point with $g_\ast >0$ and $\lambda_\ast >0$ so that gravity becomes asymptotically free in this limit.
This UV fixed point is known to exist within the Einstein-Hilbert truncation of pure gravity. The assumption is
that the coupled system of gravity plus matter behaves qualitatively in the same way.

\item[{\it (ii)}] We assume that  the system of RG improved cosmological evolution equations (3.24) 
with $k\propto 1/t$ is valid for all times $t$ after the big bang. 
This assumption means that the dominant quantum corrections 
are correctly incorporated by substituting $G_0\rightarrow G(t)$, $\Lambda_0\rightarrow\Lambda(t)$ in 
Einstein's equations and that no further modifications need to be taken into account explicitly 
(higher curvature terms, etc.). This assumption is consistent with {\it (i)} where it is also assumed that the Einstein-Hilbert
action is sufficient to describe physics for $k\rightarrow \infty$ or $t\searrow 0$.

\item[{\it (iii)}] We assume that all matter fields can be integrated out completely before solving the 
gravitational equations. This is supposed to lead to an effective 
conserved energy momentum tensor
$T_{\mu\nu}$ with a linear equation of state, $p=w\rho$. 
(However, quantum effects in the matter sector can influence
$g_\ast$ and $\lambda_\ast$, and they may shift away $w$ from its classical value.) This assumption means that, 
consistently with {\it (i)}, there are no renormalization effects coming from the matter sector which 
would be more important than those of pure quantum gravity. 
\end{enumerate}

In conclusion it is clear that cosmologies of the kind found in this paper are certainly extremely interesting
and promising candidates for an extrapolation of classical FRW cosmology 
towards earlier cosmological times and for a possible solution of its problems and limitations. 
Their most attractive
feature is that the resolution of those problems is obtained at a very low price. 
No ``ad hoc'' additional geometric
structures, matter fields or cosmological eras have to be invoked. All that is needed is the quantization of the 
fields which are present anyway.
\section{Acknowledgements}
M.R. would like to thank the Department of Theoretical Physics, University of Catania, the
Department of Theoretical Physics, University of Trieste, and the Astrophysical 
Observatory of Catania for their hospitality while this work was in progress. He also acknowledges
the financial support by INFN, MURST, and by a NATO traveling grant. 
\appendix{}
\renewcommand{\theequation}{A.\arabic{equation}}
\setcounter{equation}{0}
\section{The cutoff \normalsize{\lowercase{$k\propto 1/a $}}}
In this appendix we analyze the system of differential equations
(3.24) under the assumption that the relevant cutoff momentum is given by the inverse
scale function:
\be\label{A.1}
k(t) = {\xi\over a(t)}
\ee
Since this cutoff functionally depends on the unknown function $a(t)$, it is less straightforward
to find solutions than for the $1/t$-cutoff. We begin by solving the conservation law (\ref{3.22b})
for the density $\rho$. From (\ref{3.24}) we have 
\be\label{A.2}
\rho(t)= {{\cal M}\over 8\pi \; a(t)^ {3+3w}} 
\ee
Next we insert (\ref{A.2}) into (\ref{3.22a}) and (\ref{3.22c}) and re-express the time 
derivatives in the latter equation according to $\dot{G}= (dG/da)\dot{a}$, $\dot{\Lambda}= 
(d\Lambda/da) \dot{a}.$ Clearly this trick is possible only for cutoffs such as (\ref{A.1}) 
for which the time dependence of $k$ is purely implicit. Thus we have to solve the system
(for $\dot{a}\not = 0$): 
\begin{subequations}
\ba
&&\Big ( {\dot{a}\over {a} }\Big )^2+ {K\over a^2} = 
{\Lambda \over 3}+ {{\cal M}G \over 3 \; a^{3+3w}}\label{A.3a}\\[2mm]
&&{d\Lambda \over da}+{{\cal M}\over a^{3+3w}}\; {dG\over da} = 0\label{A.3b}\\[2mm]
&&G(t) = G(k = \xi/a), \;\; \Lambda(t) = \Lambda(k=\xi/a)\label{A.3c}
\ea
\end{subequations}
It is interesting that Eq.(\ref{A.3b}) can be rewritten directly in terms of the RG beta-functions:
\be\label{A.4}
k{d\Lambda\over d k} +{\cal M}\; \Big ( {k\over \xi}\Big )^{3+3w}\;
k{d G\over dk} = 0 
\ee
Let us look at the fixed point regime and the perturbative regime separately.
\subsection{The fixed point regime}
In the fixed point regime (\ref{A.3c}) assumes the form 
\be\label{A.5}
G(t)=\widetilde{g}_\ast \; a^2, \;\;\;\; \Lambda(t) = \widetilde{\lambda}_\ast \;a^{-2}
\ee
Again we set 
\be\label{A.cinque}
\widetilde{g}_\ast \equiv g_\ast \;\xi^{-2}, \;\;\;\; 
\widetilde{\lambda}_\ast\equiv \lambda_\ast \; \xi^2
\ee
but the constant $\xi$ differs from the one occurring in the $1/t$-cutoff. If we now
insert (\ref{A.5}) into (\ref{A.3b}) we find that this equation is satisfied provided
\be\label{A.6}
w = + {1\over 3} \;\;\text{and} \;\; \widetilde{\lambda}_\ast = {\cal M}\; \widetilde{g}_\ast
\ee
A consistent solution can be obtained 
only for the $w =1/3$ equation of state, satisfied by classical radiation for instance.
The second condition of (\ref{A.6}) will be used in order to determine $\xi$:
\be\label{A.7}
\xi^2 =\sqrt{ {g_\ast \; {\cal M}\over \lambda_\ast}} 
\ee
The last equation to be checked is (\ref{A.3a}). Plugging in $w = 1/3$, (\ref{A.5}) and
(\ref{A.7}) it boils down to the trivial differential equation $\dot{a}={\rm const}$
which, for the initial condition $a(0)=0$, is solved by $a\propto t$. Taking everything
together we see that for $w =+1/3$ there exists the following consistent solution
for all three cases $K=0$, $-1$, and $+1$: 
\begin{subequations}
\ba\label{A.8a}
&&a(t)=\Big [ {2\over 3}\;  \sqrt{g_\ast \lambda_\ast{\cal M}}-K\Big ]^{1/2} \; t \\[2mm]
&&\rho(t)={{\cal M}\over 8\pi}\; \Big [{2\over 3}\sqrt{g_\ast\lambda_\ast{\cal M}}
-K\Big ]^{-2}\; t^{-4}\label{A.8b}\\[2mm]
&&G(t)=\Big ( {g_\ast \lambda_\ast \over {\cal M}}\Big )^{1/2}
\;\Big [{2\over 3}\sqrt{g_\ast \lambda_\ast {\cal M}} -K\Big ] \; t^2\label{A.8c}\\[2mm]
&&\Lambda(t)=\sqrt{g_\ast\lambda_\ast{\cal M}}\;\Big 
[{2\over 3}\sqrt{g_\ast \lambda_\ast {\cal M}}-K\Big ]^{-1}\; t^{-2}
\label{A8.d}
\ea
\end{subequations}
We observe that (A.9) coincides precisely with (4.27) for
$K=0$ and with (4.30) derived for $K=\pm 1$. Contrary to the situation with the $1/t$-cutoff, 
no solution exists for $w \not = 1/3$, not even if $K=0$. 
\subsection{The perturbative regime}
In the perturbative regime we have 
\ba\label{A.9}
&&G(t) = G_0-\widetilde{\omega} \; G_0^2 \; a^{-2} + \cdots\\
&&\Lambda(t) =\Lambda_0 +\tilde{\nu} \; G_0 \;a^{-4} +\cdots\nonumber
\ea 
with $\widetilde{\omega}\equiv \omega\xi^2$ and $\widetilde{\nu}\equiv \nu\xi^4$. By using 
(\ref{A.9}) in (\ref{A.3b}) the following conditions arise:
\be\label{A.10}
w = - {1\over 3} \;\;\;\text{and} \;\;\; 2\;\widetilde{\nu} = \widetilde{\omega}\; {\cal M}\;  G_0
\ee
Consistency can be achieved only for the rather exotic matter with $w = -1/3$ but not for the physically
relevant cases with $w= +1/3$ or $w=0$, for instance. If we insert (\ref{A.9}) and (\ref{A.10}) into
(\ref{A.3a}) we obtain the differential equation which determines $a(t)$:
\be\label{A.11}
\dot{a}^2 +K = {1\over 3}\Lambda_0 a^2
+{1\over 3} {\cal M} G_0 -{1\over 6} \; \widetilde{\omega}\; {\cal M} \; G_0^2 \; a^{-2}+\cdots
\ee
To lowest order in $1/a$, the solution to this equation is precisely the classical FRW solution for $w=-1/3$.

To summarize: In the fixed point regime and for $w=+1/3$ the $1/a$-cutoff leads to precisely the
same cosmology as the $1/t$-cutoff. For $w\not = +1/3$ there are no solutions in the fixed point
regime. In the perturbative regime solutions exist only for
the exotic equation of state with $w=-1/3$. Because the fixed point regime and the perturbative regime
describe the limiting cases of $t\searrow 0$ and $t\rightarrow \infty$, respectively, we must conclude that,
at least with the (perhaps too poor) approximations we used, there exists no solution with constant $w$,
valid from $t=0$ up to the beginning of the classical era, which would connect to a standard radiation
dominated FRW cosmology.
\renewcommand{\theequation}{B.\arabic{equation}}
\setcounter{equation}{0}
\section{RG improvement of the classical FRW solutions}
In the main body of the paper we made the improvement $G_0\rightarrow G(t)$, 
$\Lambda_0\rightarrow \Lambda(t)$ in the {\it equations} which determine the time
evolution of $a(t)$ and the other quantities of cosmological interest. In this appendix we discuss an 
alternative strategy: the improvement of the {\it solutions} to the classical equations. In this second
approach one first solves the differential equations containing $G_0$ and $\Lambda_0$, and then one
makes the replacements $G_0\rightarrow G(t)$, $\Lambda_0\rightarrow \Lambda(t)$ in their solutions.
If $k(t)$ has an 
implicit time dependence, $G(t)$ and $\Lambda(t)$ will depend on the classical solution 
$a_{\rm class}(t)$  through $k=k(t,a_{\rm class}(t), \dot{a}_{\rm class}(t), \cdots)$. It seems clear,
and we shall demonstrate this in detail, that the method of improving equations is superior to the
improvement of solutions. In the latter case only small quantum corrections which do not change the 
behavior of the solution too strongly can be dealt with reliably, while with the first method also 
solutions which are qualitatively different from the classical ones can be investigated.

The starting point is the classical Friedmann equation 
\be\label{B.1}
\Big ( {\dot{a}\over a} \Big )^2 +{K\over a^2}= 
{\Lambda_0 \over 3} +{{\cal M} G_0 \over 3 a^{3+3w}}
\ee
from which $\rho$ has been eliminated via the conservation law (\ref{3.24}), 
\be\label{B.2}
\rho\;  = \; {{\cal M}\over 8\pi \; a^{3+3w}}
\ee
We restrict our analysis to the case $K=0$ for which the solutions to (\ref{B.1}) can be expressed 
in terms of elementary functions. Omitting the subscript ``class'', they read
(as always, for the initial condition $a(0)=0$):

\begin{enumerate}
\item[{\it (i)}] For $K=0$, $\Lambda_0=0$:
\be\label{B.3}
a(t) = \Big [ {3\over 4} (1+w)^2 {\cal M} G_0 \Big ]^{1/(3+3w)} \; t^{2/(3+3w)}
\ee
Hence, for any $w$,
\be\label{B.4}
\rho (t) = {1\over 6 \pi (1+w)^2 \; G_0 \; t^2}
\ee
\item[{\it (ii)}] For $K=0$, $\Lambda_0>0$:
\be\label{B.5}
a(t) = 
\Big [ {{\cal M} G_0 \over 2 \Lambda_0} \Big \{ {\rm cosh}[(1+w)\sqrt{3\Lambda_0} \; 
t]-1\Big \} \Big ]^{1/(3+3w)}
\ee
We shall need the Taylor expansion of this scale factor for early times $t \ll 1 /\sqrt{\Lambda_0}$:
\be\label{B.6}
a(t) = \Big [ {3\over 4} (1+w)^2 {\cal M} G_0 \; t^2\Big ]^{1/(3+3w)}\Big \{ 1+{1+w\over 12}\Lambda_0 t^2+
O(\Lambda_0^2 t^4)\Big \}
\ee
\item[{\it (iii)}] For $K=0$, $\Lambda_0 < 0$:
\be\label{B.7}
a(t) = 
\Big [{{\cal M} G_0 \over 2 |\Lambda_0|} \Big \{ 1-{\rm cos}[(1+w)\sqrt{3|\Lambda_0|} \; t]\Big \} 
\Big ]^{1/(3+3w)}
\ee
\end{enumerate}

Next we shall discuss the improvement of these solutions in the perturbative and in the 
fixed point regime, respectively. We use the identification $k=\xi /t$ throughout.
\subsection{The perturbative regime}
In this regime, $t$ may be close to the Planck time so that quantum effects are important, but
it is assumed that the lowest order terms in the $\tp/t$-expansion are sufficient to describe them:
$t\gtrsim \tp$. Furthermore we assume that, as in the real Universe, $\Lambda_0$ is small:
$\Lambda_0 \ll \mp^2$. The epoch we are interested in is characterized by 
\footnote{For definiteness we assume that $\Lambda_0>0$.}
\be\label{B.8}
\tp \lesssim t \ll 1/\sqrt{\Lambda_0}
\ee
This interval contains the late part of the Planck era where quantum gravity still 
plays a role, as well as the classical
era before the effect of the cosmological constant becomes dominant. During this epoch the product $\Lambda_0 t^2$
is small so that it is legitimate to base the improvement on the expanded form of the classical 
solution, Eq.(B.6).
Thus the RG improved scale factor reads
\be\label{B.9}
a_{\rm imp}(t) = \Big [{3\over 4} (1+w)^2{\cal M}\;  t^2\Big ]^{1/(3+3w)}
\; \Big [G(t) \Big ]^{1/(3+3w)}\; \Big \{1+{1+w\over 12}\Lambda(t) t^2+\cdots\Big \}
\ee
Using the $1/t$-cutoff, $G(t)$ and $\Lambda(t)$ are given by Eqs.(\ref{3.7}) and (\ref{3.8}), 
respectively. Hence we find the result
\be\label{B.10}
a_{\rm imp}(t) = \Big [ {3\over 4} (1+w)^2 {\cal M} G_0 \; t^2 \Big ]^{1/(3+3w)}
\Big \{ 1+{1+w\over 12}\Lambda_0 t^2 
+ \Big ( {(1+w)\widetilde{\nu}\over 12}-{\widetilde{\omega}\over 3 (1+w)}\Big )
\Big ({\tp \over t}\Big )^2 +\cdots \Big \}
\ee
The leading quantum correction is a modification of $a(t)$ by a term of order
$(\tp / t)^2$. Within the present approach, its prefactor is completely undetermined, however.
It involves the parameter $\xi$ which cannot be fixed by renormalization group arguments
alone. The method of improving equations is much more powerful in this respect; 
it allows us to express 
$\xi$ in terms of $\omega$ and $\nu$. In a kind of hybrid calculation we could use this result
in order to rewrite (\ref{B.10}). This would change the terms inside the curly brackets of 
(\ref{B.10}) to 
\be\label{B.11}
1-{5\omega^2\over 27 \nu (1+w)^3 }\Big ( {\tp \over t}\Big )^2+\cdots
\ee
(We also took the other one of the consistency conditions (\ref{4.6}), 
$\Lambda_0 = 0$, into account.)

It is important to note that a correction term of the type (\ref{B.11}) could not have been 
found as a solution to the improved {\it equation} unless one includes in $G(t)$ and 
$\Lambda(t)$ higher orders of the $\tp/t$-expansion. The reason is the remarkable fact, discussed 
in Section 4, that the classical $a(t)$ arises as a consequence of the lowest order nontrivial 
time dependence in $G(t)$ and $\Lambda(t)$.
\subsection{The fixed point regime}
Let us  look at the improvement for $t\ll \tp$. In this regime the renormalization effects are strong 
and strictly speaking it is not clear if the results are reliable. We start from the classical $K=0$, 
$\Lambda_0>0$ solution (\ref{B.5}) and substitute $G_0\rightarrow G(t)$, $\Lambda_0 \rightarrow \Lambda(t)$
according to Eq.(\ref{3.10}) and Eq.(\ref{3.11}), respectively. This substitution turns the ${\rm cosh}(t)$-time
dependence into a purely algebraic one:
\be\label{B.12}
a_{\rm imp}(t) = 
\Big [ {\widetilde{g}_{\ast}
{\cal M}\over 2\widetilde{\lambda}_\ast} \Big\{ {\rm cosh}[(1+w)\sqrt{3\widetilde{\lambda}_\ast}]
-1\Big\}\Big]^{1/(3+3w)} \; t^{4/(3+3w)}
\ee
It is reassuring that apart from the details of the prefactor (\ref{B.12}) coincides with our previous
result obtained by improving equations, Eq.(\ref{4.20a}). 

To summarize: Improving the classical FRW solutions shows that for $t\searrow 0$ the onset of
the Planck era is characterized by a $(\tp/t)^2$-correction to the scale factor. In the fixed point 
regime, this approach provides an independent confirmation of the picture we obtained by RG improving the 
equations.

\end{document}